\documentclass[journal]{IEEEtran}
\usepackage{cite}
\usepackage{bigints}
\usepackage{amsmath}
\usepackage{array}
\usepackage[pdftex]{graphicx}
\usepackage{xcolor}
\usepackage{bm} 
\usepackage{amssymb}
\usepackage{dsfont}
\usepackage{hyperref}

\allowdisplaybreaks    

\def\Xint#1{\mathchoice
   {\XXint\displaystyle\textstyle{#1}}%
   {\XXint\textstyle\scriptstyle{#1}}%
   {\XXint\scriptstyle\scriptscriptstyle{#1}}%
   {\XXint\scriptscriptstyle\scriptscriptstyle{#1}}%
   \!\int}
\def\XXint#1#2#3{{\setbox0=\hbox{$#1{#2#3}{\int}$}
     \vcenter{\hbox{$#2#3$}}\kern-.5\wd0}}

\def\dashint{\Xint-}

\newcommand{\rpos}{\mathbf{r}}
\newcommand{\rposp}{\mathbf{r}'}

\title{All-analytical evaluation of the singular integrals involved in the Method of Moments}

\author{%
D. Tihon \textit{Student Member, IEEE}
and C. Craeye \textit{Senior Member, IEEE}
\thanks{Universit\'{e} Catholique de Louvain, ICTEAM institute, Place du Levant 3, 1348 Louvain-la-Neuve, Belgium (e-mails: denis.tihon@uclouvain.be, christophe.craeye@uclouvain.be).}}

\begin{document}

\maketitle

\begin{abstract}
Surface Integral Equation (SIE) methods routinely require the integration of the singular Green's function or its gradient over Basis Functions (BF) and Testing Functions (TF). Many techniques have been described in the literature for the fast and accurate computation of these integrals for TF that is located close to the BF. In this paper, we propose an all-analytical formula for the singular part of the integral for both the Electric and Magnetic Field Integral Equations (EFIE and MFIE). The method works for any flat polygonal BF and TF of any order, and proves to be competitive with existing techniques.
\end{abstract}

\begin{IEEEkeywords}
Singular integral, Integral Equation Method, Method of Moments.
{\color{blue} \textcopyright 2018 IEEE.  Personal use of this material is permitted.  Permission from IEEE must be obtained for all other uses, in any current or future media, including reprinting/republishing this material for advertising or promotional purposes, creating new collective works, for resale or redistribution to servers or lists, or reuse of any copyrighted component of this work in other works. The published version is available at \url{https://doi.org/10.1109/TAP.2018.2803130}.}
\end{IEEEkeywords}

\section{Introduction}
In computational electromagnetics, Surface Integral Equations (SIE) based methods are popular since only the surfaces need to be discretized, leading to a limited number of unknowns. These methods also offer a great flexibility for physics-based acceleration techniques (see e.g: fast multipoles \cite{FMM}, macro-basis functions \cite{MBF}, ...). However, such techniques require to efficiently evaluate singular ($1/R$) and hypersingular ($1/R^2$) 4-dimensional (4D) integrals over both Basis Functions (BF) and Testing Functions (TF), which appear in the Electric Field Integral Equation (EFIE) and Magnetic Field Integral Equation (MFIE), respectively. The evaluation of these integrals can be challenging if the TF is located close to the BF ($R \rightarrow 0$).

In the literature, many techniques have been proposed to speed up the evaluation of these integrals. They can mainly be classified into two categories: singularity extraction based techniques \cite{Wilton84, Eibert95dec, YlaOijala2003, Jarvenpaa2003, Jarvenpaa2006, Graglia1993, Hodge1996, Tong2007, Tong2009, Polimeridis2010march, Polimeridis2011, Polimeridis2008sept, Polimeridis2011june, Gurel2005, Bleszynski2016, Arcioni1997march, Caorsi1993sept}, and singularity cancellation based techniques \cite{Ismatullah2008, Li2016, Botha2015, Botha2013, Vipiana2007, Vipiana2008june, Vipiana2013march, Vipiana2013November, Polimeridis2010june, Polimeridis2013june, Kaur2011, Khayat2005, Khayat2008, Fink2013, Yuan2015june, Graglia2008april, Polimeridis2013}.

The singularity extraction techniques rely on the analytical evaluation of the part of the integral that is singular and the numerical computation of the remaining part, which is either a integral of reduced dimensionality or a 4D integral whose integrand is bounded. In \cite{Wilton84}, the authors propose an analytical formula for the integration of the static kernel of the Green's function over the BF. As suggested in \cite{Arcioni1997march}, the remaining 2D integral over the TF is sufficiently smooth to be integrated numerically using Gaussian quadrature schemes. In \cite{Eibert95dec}, an analytical formula is even proposed for the full 4D integral of the static Green's function over coincident BF and TF. Once the static kernel have been removed from the dynamic Green's function, the integrand of the remaining 4D integral is bounded. However, its derivatives are not, so that its numerical integration may remain challenging. In \cite{Caorsi1993sept, YlaOijala2003}, the convergence of the numerical integral is further accelerated by extracting additional terms analytically. 

The analytical integral of the MFIE term over the BF has been proposed in \cite{Graglia1993, Hodge1996} for linear BF, and extended to higher orders BF in \cite{Jarvenpaa2003, Jarvenpaa2006}. However, the remaining integral over the TF includes a logarithmic singularity that has to be handled carefully if the BF and TF share a common point \cite{YlaOijala2003, Gurel2005}. The technique has even been extended to the super-hypersingular kernel ($1/R^3$) that may arise in some formulation of the EFIE \cite{Tong2007, Tong2009}.
More recently, techniques in which the two 2D integrals over the BF and TF are decomposed into four 1D integrals whose order can be switched has been proposed, leading to even better results \cite{Bleszynski2016, Polimeridis2010march, Polimeridis2011, Polimeridis2008sept, Polimeridis2011june}.

The singularity cancellation based techniques are purely numerical techniques. Generally, an irregular sampling of the integrand is used in order to minimize the number of sampling points required to reach a prescribed accuracy. The problem is formulated as a change of the variables of integration. The new variables are chosen such that the Jacobian of the transformation compensates for the singular behaviour of the integrand, so that the new integrand is smooth and can be integrated using classical quadrature schemes. 

An elegant alternative technique has been proposed in \cite{Vipiana2007, Vipiana2008june}. The authors use a non-singular approximate of the Green's function, which corresponds to the exact Green's function that has been filtered in spectral domain. The error introduced when filtering the Green's function is mitigated by the mesh used to discretize the geometry, which naturally acts as a low-pass filter.

In this paper, we propose an approach to solve analytically the 4D singular integrals that appear in EFIE and MFIE formulations. The method we propose can be used to solve the 4D integral of the static kernel of the Green's function and its gradient, but also the higher order terms \cite{YlaOijala2003, Caorsi1993sept}. Instead of decomposing the integral over the BF and TF into a sum of 2D or 1D integrals, the initial 4D integral is treated as a whole. The proposed formulation starts by identifying a 4D volume of integration in a 6D space, on which the distance function only varies over three of the six dimensions. By using the divergence theorem, the 4D integral can be reduced to 3D, 2D and then 1D integrals, the latter being solved analytically. Note that the framework of the proposed method is also valid for integral over 3D BF and TF. Therefore the proposed method can be straightforwardly extended to volumetric MoM. Preliminary results for the static EFIE and non-parallel BF and TF have been presented in \cite{Tihon2017}.

In Section \ref{sec:problem_definition}, the different types of integrals that may arise in the EFIE and MFIE for polynomial BF and TF are highlighted. Then, in Section \ref{sec:6Dgeom}, the 6D geometry in which the integration takes place is described. In Section \ref{sec:6Dto3D}, the 4D integral taking place in the 6D space is reduced to a sum of classical 3D or 2D integrals taking place in the 3D space. The latter integrals are solved in Section \ref{sec:3Dint}. Section \ref{sec:Ci} is dedicated to the treatment of the hypersingular integral that appears in the MFIE. Finally, the technique is validated through representative examples in Section \ref{sec:numerical_validation}.

\section{Problem definition}
\label{sec:problem_definition}
Consider the interaction between two flat polygonal BF and TF that are div-conforming. The BF and TF vector fields are polynomial functions of the coordinates, $\mathbf{F}_B(x,y,z)$ and $\mathbf{F}_T(x,y,z)$. Without losing generality, we will chose a right-handed coordinate system such that $\hat{z}$ corresponds to the normal to the BF $\hat{n}_B$, $\hat{x}$ is the direction that is parallel to both the BF and TF $\hat{x} = (\hat{n}_B \times \hat{n}_T)/|\hat{n}_B \times \hat{n}_T|$, with $\hat{n}_T$ the normal to the TF, and the $\hat{y}$ direction is chosen accordingly. If the BF and TF are parallel, then $\hat{x}$ is any direction parallel to the BF. The impedance matrix entries for a pair of div-conforming BF and TF are given by \cite{MoM}
\begin{subequations}
\label{eq:25-01-01}
\begin{align}
Z^{EJ} &= \dfrac{j\eta}{4\pi k} \int_{S'} \int_{S} \Big( k^2 \mathbf{F}_B(\mathbf{r}) \cdot \mathbf{F}_T(\mathbf{r'}) \\
&~~~~ -\big(\mathbf{\nabla} \cdot \mathbf{F}_B(\mathbf{r})\big) \big(\mathbf{\nabla}' \cdot \mathbf{F}_T(\mathbf{r'})\big) \Big) \dfrac{e^{-jkR}}{R} dS'(\mathbf{r'}) dS(\mathbf{r}), \nonumber\\
Z^{EM} &= \dfrac{1}{4\pi} \dashint_{S'} \int_{S}  \mathbf{\nabla}' \bigg(\dfrac{e^{-jkR}}{R}\bigg) \label{eq:22-02-01}\\
&\nonumber ~~~~~~~~~~~~~~~~~~~~~ \times \mathbf{F}_B(\mathbf{r}) \cdot \mathbf{F}_T(\mathbf{r'}) dS'(\mathbf{r'}) dS(\mathbf{r}), 
\end{align}
\end{subequations}
with $Z^{EJ}$ and $Z^{EM}$ the electric field generated on the TF by electric ($J$) and magnetic ($M$) currents flowing on the BF, $\mathbf{r}=(x,y,z)$ and $\mathbf{r}'=(x',y',z')$ the position in the BF and TF respectively, $\mathbf{\nabla}$ and $\mathbf{\nabla}'$ the derivative operators with respect to coordinates $\mathbf{r}$ and $\mathbf{r}'$, respectively, $R = |\mathbf{R}|$ the distance function, $\mathbf{R} = \mathbf{r}'-\mathbf{r}$, $k$ and $\eta$ the wavenumber and impedance corresponding to the medium through which the BF and TF are interacting, $S$ and $S'$ the surfaces of the BF and TF, respectively, and $\dashint$ the integral in the Cauchy principal value meaning. Note that the overlap integral have been removed from (\ref{eq:22-02-01}) since the Poggio-Miller-Chew-Harrington-Wu-Tsai (PMCHWT) formulation of the MoM is considered in \cite{MoM}. However, this integral can be easily computed numerically if required \cite{Jarvenpaa2006} and won't be treated hereafter.

The exponential term in the integrands of (\ref{eq:25-01-01}) can be expanded into a Taylor series of order $N$ around the value $R_o$ \cite{FMIR}, leading to
\begin{equation}
\label{eq:25-01-02}
\begin{split}
e^{-jkR} &= e^{-jkR_o} e^{-jk(R-R_o)} \\
&\simeq e^{-jkR_o} \sum_{i=0}^{N} \dfrac{\big(-jk(R-R_o)\big)^i}{i!} \\
&= e^{-jkR_o} \sum_{i=0}^{N} K_i^N(k, R_o) R^i.
\end{split}
\end{equation}
with 
\begin{equation}
K_i^N(k, R_o) = \sum_{\alpha=i}^{N} \dfrac{(-j k) ^\alpha}{\alpha!} \begin{pmatrix} \alpha \\ i\end{pmatrix}  (-R_0)^{\alpha-i}.
\end{equation}

Substituting (\ref{eq:25-01-02}) into (\ref{eq:25-01-01}), using the linearity of the derivative operator and isolating the wavenumber $k$ gives
\begin{subequations}
\label{eq:04-04-02}
\begin{align}
Z^{EJ} &= \dfrac{j\eta\exp(-jkR_0) }{4\pi k} \sum_{i=0}^N K_i^N(k,R_o) (k^{2} A_i -  B_i),
\\
Z^{EM} &= \dfrac{\exp(-jkR_0)}{4\pi} \sum_{i=0}^N K_i^N(k,R_o) C_i
\end{align}
\end{subequations}
with the frequency and material independent terms \cite{FMIR}
\begin{subequations}
\label{eq:04-04-01}
\begin{align}
&A_i = \int_{S'} \int_{S} R^{(i-1)} \big(\mathbf{F}_B \cdot \mathbf{F}_T \big) dS~ dS',\\
&B_i = \int_{S'} \int_{S} R^{(i-1)} \big(\mathbf{\nabla} \cdot \mathbf{F}_B \big) \big(\mathbf{\nabla}' \cdot \mathbf{F}_T \big) dS~ dS' ,\\
&C_i = (i-1) \dashint_{S'} \int_{S} R^{(i-3)} \mathbf{R} \times \mathbf{F}_B \cdot \mathbf{F}_T dS~ dS' .
\end{align}
\end{subequations}
The $\mathbf{r}$ and $\mathbf{r}'$ dependencies have been omitted for clarity. 
This formulation is very convenient since each term of (\ref{eq:04-04-01}) is independent from frequency and material parameters, so that they can be used for any frequency and materials, as long as the geometry does not change. Moreover, since the Taylor expansion (\ref{eq:25-01-02}) is applied to the phase term only, the convergence of the series is very rapid (Taylor expansion of a complex exponential). Last, if $R_0$ is chosen to correspond to the mean distance between the BF and TF, the convergence of (\ref{eq:04-04-01}) is independent from the distance between the BF and TF. An extensive investigation of the number of terms required to reach a prescribed level of accuracy has already been carried out in \cite{FMIR} and is out of the scope of the present paper.

It can be noticed that the three sets of integrals in (\ref{eq:04-04-01}) can be reduced to two sets of canonical integrals.
As mentioned in the beginning of this Section, the vector fields of the BF and TF can be expressed as polynomials of the coordinates, so that the generic shapes of $A_i$ and $B_i$ and of $C_i$, respectively $\mathcal{A}_{i}$ and $\mathcal{C}_i$, read:
\begin{subequations}
\label{eq:25-01-03}
\begin{align}
\mathcal{A}_i &=\int_{S'} \int_{S} R^{(i-1)} \mathcal{P}_1(\mathbf{r}, \mathbf{r}') dS~ dS' \label{eq:Ai},\\
\mathcal{C}_i &= (i-1) \dashint_{S'} \int_{S} R^{(i-3)} \big(\mathbf{R} \cdot \bm{\mathcal{P}}_1(\mathbf{r}, \mathbf{r}') \big) dS~ dS' \label{eq:Ci},
\end{align}
\end{subequations}
with $\mathcal{P}_k(\mathbf{r}, \mathbf{r}')$ and $\bm{\mathcal{P}}_k(\mathbf{r}, \mathbf{r}')$ corresponding to polynomial scalar and vector functions of the positions $\mathbf{r}$ and $\mathbf{r}'$. 
The evaluation of the terms of (\ref{eq:25-01-03}) may be difficult to handle numerically, since the integrand and/or its higher derivatives are not bounded for even $i$ \cite{Caorsi1993sept,YlaOijala2003}, requiring semi-analytical techniques or advanced quadrature schemes. The goal of this paper is to solve analytically these integrals. 

Most of the time, in the literature, $R = R(\rpos-\rposp)$ is seen as a 3D function $R: \mathds{R}^3 \rightarrow \mathds{R}: \rpos \rightarrow |\rpos|$. Due to this, the two surface integrals of (\ref{eq:25-01-03}) are difficult to merge into a single 4D integral since this 4D integral would take place in a 3D space. For this reason, the starting point of this paper is to consider that $R = R(\rpos, \rposp)$ is a 6D function $R: \mathds{R}^6 \rightarrow \mathds{R} : (\rpos, \rposp) \rightarrow |\rpos-\rposp|$. The advantage of this formulation is that, now, the $\rpos$ and $\rposp$ vectors can be seen as two orthogonal vectors in the 6D space, $(\rpos, \mathbf{0})$ and $(\mathbf{0}, \rposp)$ respectively, such that the two consecutive 2D integrals of (\ref{eq:25-01-03}), which both take place in a 3D space, can be merged into a single 4D integral in the 6D space.

Note that the $\mathcal{C}_i$ term is a particular case of the $\mathcal{A}_{i-1}$ term, but with additional properties that will prove to be useful to treat the singular case $R\rightarrow 0$. The method used for both terms being essentially identical, the treatment of the $\mathcal{A}_i$ term will be used as an illustration, while the specificities in the treatment of the $\mathcal{C}_i$ term are provided in Section \ref{sec:Ci}.

\section{The 6D geometry}
\label{sec:6Dgeom}
Each term of (\ref{eq:25-01-03}) corresponds to two consecutive surface integrals (2D) over BF and TF, which lie in the 3D space. Therefore, it can be reformulated in the sense of a 4D integral that takes place within a 6D space, each point $(\mathbf{r}, \mathbf{r}')$ of the 6D space corresponding a pair of points, one on the BF and one on the TF, of coordinates $\mathbf{r}$ and $\mathbf{r}'$, respectively. Then, the 4D volume of integration within this 6D space corresponds to the combination of the BF with the TF in the 6D space.

\begin{figure}[h!]
\center
\includegraphics[width = 9cm]{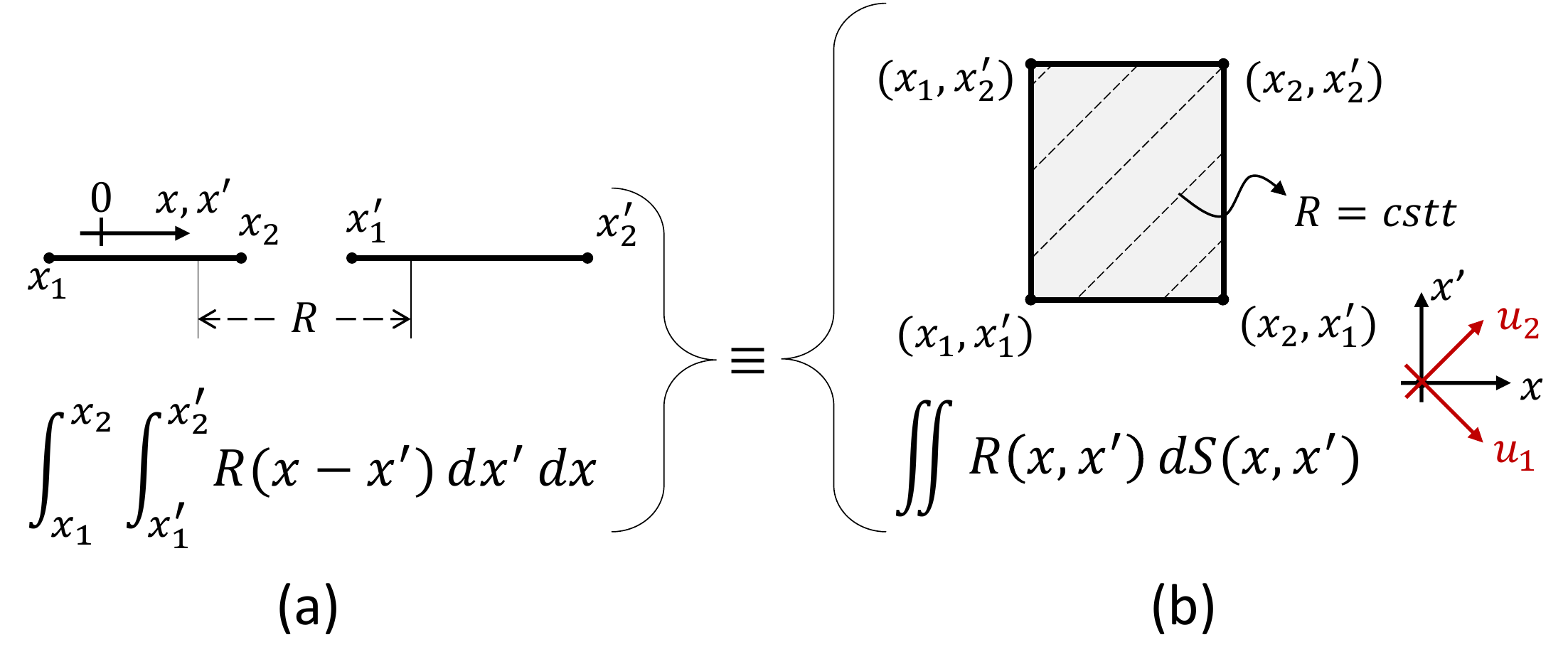}
\caption{The two 1D integrals of (a) are merged into a single 2D integral over the gray surface of (b), which can be solved more easily. First, the divergence theorem can be used. Second, rotating the system of coordinates from $\{x,x'\}$ to $\{u_1, u_2\}$, the $R$ function is independent from the $u_2$ coordinate.}
\label{fig:1Dto2D}
\end{figure}

To begin with, let us illustrate this manipulation with a 1D example. Consider the integration of the distance function $R$ over a 1D BF and a 1D TF within a 1D space, as illustrated in Fig. \ref{fig:1Dto2D}(a). If $x$ denotes the coordinate of the points in the BF, and $x'$ denotes the coordinate of the points in the TF, the distance function becomes $R(x-x') = |x - x'|$. At first glance, the two 1D integrals must be solved sequentially. 

Re-writing the problem in two dimensions, one obtains Fig. \ref{fig:1Dto2D}(b). Since $x$ and $x'$ are independent, they correspond to two independent directions. In this new 2D space, the two 1D integrals become a single surface (2D) integral. This reformulation has two main advantages. First, the divergence theorem can be applied to the surface of Fig. \ref{fig:1Dto2D}(b) to reduce the 2D integral to a sum of 1D integrals. Second, it can be seen that the $R$ function is very easy to deal with using a change of coordinates.  Indeed, using the $\{u_1, u_2\}$ system of coordinates indicated in Fig. \ref{fig:1Dto2D}(b), the $R$ function only varies along the $\hat{u}_1$ direction and is constant along the $\hat{u}_2$ direction. Similar reasoning will prove to be very useful for the analytical solution of (\ref{eq:25-01-03}).

Going back to the original problem, since the whole technique is based on the manipulation of the 4D volume in 6D space, it is worth spending some times to clarify the geometry considered.

First of all, consider the 4D volume itself. Since the integration takes place in a 6D space, the 4D space spanned by the 4D volume is characterized by two normal directions, $\hat{n}_{v,1}^6$ and $\hat{n}_{v,2}^6$ (just as a line is in the 3D space). These unit vectors are corresponding to the normals to the BF and TF, respectively $\hat{n}_B$ and $\hat{n}_T$. It is important to notice that the scalar product of a normal with the position of any point within the volume is constant. Since the BF and TF are polygonal, the 3D boundary of the 4D volume corresponds to 3D polyhedra, just as the 2D boundary of a 3D polyhedron corresponds to polygons. 

Each 3D polyhedron corresponds to the combination of one edge of the BF with the whole TF or one edge of the TF with the whole BF. In order to characterize the 3D space spanned by these volumes, one additional normal direction is required. For volume $p$, this normal will be noted $\hat{n}_p^6$. Note that many volumes $p$ are contiguous and therefore share a common surface.

The 2D boundary of one 3D polyhedron corresponds to flat polygons. They are obtained either by convolving one edge of the BF and one edge of the TF, or convolving one vertex of the BF (or TF) with the TF (or BF). The additional normal direction required to describe the 2D space spanned by the 2D surface $q$ will be noted $\hat{n}_q^6$.

The 1D edges delineating the 2D surfaces are the combination of edges of the BF (or TF) with vertices of the TF (or BF). Their corresponding normals will be noted $\hat{n}_r^6$. 

Last come the vertices of all these geometrical entities. They correspond to the combination of the vertices of the BF with the vertices of the TF, so that the coordinate of the 6D vertex corresponding to the combination of the vertex $i$ of the BF located in $\mathbf{r}_i$ and the vertex $j$ of the TF located in $\mathbf{r}'_j$ is located in $(\mathbf{r}_i, \mathbf{r}'_j)$.

Now, we consider a ``sum-and-difference" change of coordinates:
\begin{equation}
\label{eq:26-01-01}
\begin{array}{lcl}
u_1 = \dfrac{x-x'}{\sqrt{2}},
	& ~~ &
	u_4 = \dfrac{x+x'}{\sqrt{2}}, \\
u_2 = \dfrac{y-y'}{\sqrt{2}},
	& ~~ &
	u_5 = \dfrac{y+y'}{\sqrt{2}}, \\
u_3 = \dfrac{z-z'}{\sqrt{2}},
	& ~~ &
	u_6 = \dfrac{z+z'}{\sqrt{2}}.
\end{array}
\end{equation}
Using this new basis gives $\mathbf{R} = \sqrt{2}(u_1 \hat{u}_1 + u_2 \hat{u}_2 + u_3 \hat{u}_3)$, which is independent from $u_4$, $u_5$ and $u_6$. This will be fundamental for the analytical evaluation of (\ref{eq:25-01-03}). Moreover, it can be seen that, except for BF and TF defined on parallel domains ($\hat{n}_B = \hat{n}_T$), neither the $\hat{u}_5$ nor the $\hat{u}_6$ direction is contained within the space spanned by the 4D volume of integration. The case of parallel BF and TF will be treated in a dedicated Section.

In order to perform the integral over the 4D volume within the 6D space, three different tools will be repetitively used.

\begin{figure}[h!]
\center
\includegraphics[width = 4cm]{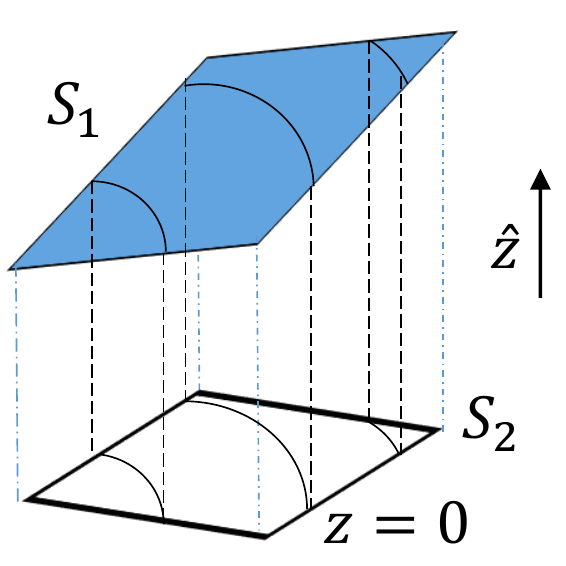}
\caption{The orthogonal projection of the surface of integration $S_1$ along the $\hat{z}$ direction does not change the value of the integral (excepted for the constant Jacobian) as long as the integrand does not depend on $z$.}
\label{fig:projection}
\end{figure}

\begin{enumerate}
\item
\emph{TOOL$\_$1:} If the integrand is modified outside the domain of integration, it does not modify the value of the integral. This can be used to create a new integrand that is independent from one chosen coordinate. For example, consider a $\hat{x}-\hat{y}-\hat{z}$ Cartesian coordinate system. The $z$ coordinate within a non-vertical 2D surface linearly depends on the two other coordinates: $z = k_1 x + k_2 y +k_3$. Thus, the integral of any function $f(x,y,z)$ within this surface can be reformulated as the integral of $g(x,y) = f(x,y, k_1 x + k_2 y +k_3)$, with $k_1$, $k_2$ and $k_3$ three constants that depend on the surface.
\item
\emph{TOOL$\_$2:} If the integrand is independent from one coordinate $u_i$, and if the corresponding direction $\hat{u}_i$ is not contained within the space spanned by the domain of integration, this domain can be projected into the subspace $u_i = 0$. The new integral will be equal to the previous one within a constant factor, which corresponds to the Jacobian of the transformation. Using this method, the dimensionality of the space in which the integration takes place is reduced. For example, consider the integral of $g(x,y)$ over the surface $S_1$ represented in Figure \ref{fig:projection}.  $S_2$ is the orthogonal projection of $S_1$ on the $z=0$ subspace. The value of the integral of $g(x,y)$ over $S_2$ is the same as its integral over $S_1$ within a constant factor corresponding to the Jacobian of the transformation.
\item
\emph{TOOL$\_$3:} If a vector field can be found, whose divergence matches the integrand within the domain of integration, applying the divergence theorem reduces the dimensionality of the domain of integration:
\begin{equation}
\begin{split}
&\text{if } \mathbf{\nabla}\cdot \mathbf{G} = f \text{ on } \Omega, \\
&\int_\Omega f ~ dV = \int_\Omega \mathbf{\nabla}\cdot \mathbf{G} ~ dV = \int_{\partial \Omega} \mathbf{G} \cdot \hat{n} ~ dS,
\end{split}
\end{equation}
with $\Omega$ the domain of integration, $\partial \Omega$ its boundary and $\hat{n}$ the outward-pointing unit normal. Note that the derivative operator $\mathbf{\nabla}$ has to be limited to the space spanned by the domain of integration. For example, if the divergence theorem is applied on a surface in the 3D space, the surface divergence must be considered instead of the volume divergence, and the surface integral can be reformulated as a line (1D) integral.
\end{enumerate}

\section{From 6D to 3D}
\label{sec:6Dto3D}
In this section, the 4D integral within 6D space of Equation (\ref{eq:Ai}) is reduced to a sum of 2D or 3D integrals within 3D space. 

Formulated in the 6D space,  (\ref{eq:Ai}) becomes
\begin{equation}
\label{eq:26-01-03}
\begin{split}
\mathcal{A}_i =\int &R^{(i-1)}(u_1, u_2, u_3) \\
&~~~~ \times \mathcal{P}_2(u_1, u_2, u_3, u_4, u_5, u_6) dV_4^6,
\end{split}
\end{equation}
with $dV_i^j$ indicating that the integration is performed over a $i$D volume within the $j$D space. 

First of all, consider the two normals to the 4D volume within the 6D space. Since these normals correspond to the normals to the BF and TF, using (\ref{eq:26-01-01}) and considering the specific definitions of the $\hat{x}$ and $\hat{z}$ directions described in Section \ref{sec:problem_definition} gives
\begin{subequations}
\label{eq:26-01-02}
\begin{align}
\hat{n}_{v,1}^6 &= (\hat{n}_B, \hat{n}_B)/\sqrt{2} = (0,0,1,0,0,1)/\sqrt{2}\\
\hat{n}_{v,2}^6 &= (-\hat{n}_T, \hat{n}_T)/\sqrt{2} = (0, -a, -b, 0, a, b)/\sqrt{2},
\label{eq:26-01-02-b}
\end{align}
\end{subequations}
with $a^2+b^2 = 1$. Note that in the case of parallel BF and TF, $a=0$. This case will be treated in Section \ref{sec:parallel}.

The next step consists in the orthogonal projection of the 4D volume into the 5D subspace $u_6=0$. First, the orthogonal vectors are combined to obtain:
\begin{subequations}
\begin{align}
\hat{n}_{v,1}^5 &= \dfrac{(\hat{n}_{v,2}^6 \cdot \hat{u}_6) \hat{n}_{v,1}^6 - (\hat{n}_{v,1}^6 \cdot \hat{u}_6) \hat{n}_{v,2}^6}{\sqrt{(\hat{n}_{v,1}^6 \cdot \hat{u}_6)^2+(\hat{n}_{v,2}^6 \cdot \hat{u}_6)^2}}  \\
&= \dfrac{b~ \hat{n}_{v,1}^6- \hat{n}_{v,2}^6}{\sqrt{1+b^2}}, \nonumber
\\
\hat{n}_{v,2}^5 &= \dfrac{(\hat{n}_{v,1}^6 \cdot \hat{u}_6) \hat{n}_{v,1}^6 + (\hat{n}_{v,2}^6 \cdot \hat{u}_6) \hat{n}_{v,2}^6}{\sqrt{(\hat{n}_{v,1}^6 \cdot \hat{u}_6)^2+(\hat{n}_{v,2}^6 \cdot \hat{u}_6)^2}}  \\
&= \dfrac{\hat{n}_{v,1}^6 + b ~ \hat{n}_{v,2}^6}{\sqrt{1+b^2}}, \nonumber
\end{align}
\end{subequations}
so that $\hat{n}_{v,1}^5 \cdot \hat{u}_6 = 0$ and $\hat{n}_{v,1} \cdot \hat{n}_{v,2} = 0$. In that way, $\hat{n}_{v,1}^5$ is orthogonal to both the original volume and its projection into the $u_6 = 0$ subspace.

\subsection{From 6D to 5D}
\label{sec:6Dto5D}
Now, the $u_6$ dependency of the integrand in (\ref{eq:26-01-03}) can be removed using \emph{TOOL$\_$1}. If $d^j_{i}$ is the distance between the volume of integration and the origin in the $\hat{n}_{i}^j$ direction, then
\begin{equation}
\hat{n}_{v,2}^5 \cdot \mathbf{u} = d^5_{v,2} ~~~~~~ \forall \mathbf{u} \in \Omega_v^6,
\end{equation}
with $\Omega_v^6$ the 4D domain of integration lying in the 6D space. Isolating the $u_6$ term provides
\begin{equation}
\label{eq:26-01-04}
u_6 = \dfrac{d_{v,2}^5 - \sum_{\alpha =1}^5 (\hat{n}_{v,2}^5 \cdot \hat{u}_\alpha) u_\alpha}{\hat{n}_{v,2}^5 \cdot \hat{u}_6}  ~~~~~~ \forall \mathbf{u} \in \Omega_v^6.
\end{equation} 
It can be easily checked that, using the particular basis described in Section \ref{sec:problem_definition}, $\hat{n}_{v,2}^5 \cdot \hat{u}_6 \neq 0$. Substituting (\ref{eq:26-01-04}) into (\ref{eq:26-01-03}) gives
\begin{equation}
\label{eq:26-01-05}
\begin{split}
\mathcal{A}_i =\int &R^{(i-1)}(u_1, u_2, u_3) \\
&~~~~ \times \mathcal{P}_3(u_1, u_2, u_3, u_4, u_5) dV_4^6.
\end{split}
\end{equation}

Now that the integrand of (\ref{eq:26-01-05}) is independent from $u_6$, an orthogonal projection of the volume of integration over the $u_6 = 0$ subspace is performed (\emph{TOOL$\_$2}) in order to reduce the dimensionality of the space we are working on.
Multiplying the resulting integral by the (constant) Jacobian of the projection provides
\begin{equation}
\label{eq:26-01-06}
\begin{split}
\mathcal{A}_i = \dfrac{1}{|\hat{n}_{v,2}^5 \cdot \hat{u}_6|} \int &R^{(i-1)}(u_1, u_2, u_3) \\
&~~~~ \times \mathcal{P}_3(u_1, u_2, u_3, u_4, u_5) dV_4^5.
\end{split}
\end{equation}

As mentioned previously, since $\hat{n}_{v,1}^5 \cdot \hat{u}_6 = 0$, $\hat{n}_{v,1}^5$ is orthogonal to both the original volume and the projected one. The projected volume being a 4D volume within a 5D space, its orthogonal space is 1D and is therefore entirely spanned by the $\hat{n}_{v,1}^5$.

\subsection{From 5D to 4D}
\label{sec:5Dto4D}
This Section is treating the case where the BF and TF are not parallel. The parallel case will be treated in Section \ref{sec:parallel}. If the BF and TF are not parallel, it means that $a \neq 0$ in (\ref{eq:26-01-02-b}). Consequently, $\hat{n}_{v,1}^5 \cdot \hat{u}_5 \neq 0$, so that the procedure to reduce the dimensionality of the space is identical when going from 6D to 5D and from 5D to 4D.

First, within the volume of integration, the $u_5$ coordinate can be expressed as a linear combination of the other coordinates (\emph{TOOL$\_$1}):
\begin{equation}
\label{eq:26-01-07}
u_5 = \dfrac{d_{v,1}^5 - \sum_{\alpha=1}^4 (\hat{n}_{v,1}^5 \cdot \hat{u}_\alpha) u_\alpha}{\hat{n}_{v,1}^5 \cdot \hat{u}_5}.
\end{equation}
Substituting (\ref{eq:26-01-07}) into (\ref{eq:26-01-06}), the $u_5$ dependence of the integrand is removed. Putting the $1/|\hat{n}_{v,2}^5 \cdot \hat{u}_6|$ factor of (\ref{eq:26-01-06}) inside the polynomial $\mathcal{P}_4$ gives:
\begin{equation}
\label{eq:26-01-08}
\mathcal{A}_i = \int R^{(i-1)}(u_1, u_2, u_3)  \mathcal{P}_4(u_1, u_2, u_3, u_4) ~ dV_4^5.
\end{equation}

Since the integrand of (\ref{eq:26-01-08}) no longer depends on $u_5$, the 5D space is projected into the 4D subspace $u_5 = 0$ (\emph{TOOL$\_$2}), leading to 
\begin{equation}
\label{eq:26-01-09}
\begin{split}
\mathcal{A}_i = \dfrac{1}{|\hat{n}_{v,1}^5 \cdot \hat{u}_5|}\int &R^{(i-1)}(u_1, u_2, u_3) \\
&~~~~\times \mathcal{P}_4(u_1, u_2, u_3, u_4) ~dV_4^4.
\end{split}
\end{equation}

It is interesting to consider the behaviour of the method for the case where the BF and TF are close to parallel, without being exactly so. Indeed, in that case, $\hat{n}_{v,1}^5 \cdot \hat{u}_5 \ll 1$.The RHS of (\ref{eq:26-01-07}) then corresponds to a nearly canceling difference between large numbers. Therefore, for very small angles (smaller than $1^\circ$), one may consider interpolating the value of the integral from its value for parallel BF and TF and for non-parallel BF and TF with a higher angle between them (e.g. $1^\circ, 2^\circ, ...$). An example using polynomial interpolation of various orders is studied in Section \ref{sec:numerical_validation}, where the proposed fix is illustrated.

\subsection{From 4D to 3D}
\label{sec:4Dto3D}
Equation (\ref{eq:26-01-09}) is dealing with a 4D volume of integration within a 4D space. The dimensionality of the volume of integration needs to be reduced using the divergence theorem (\emph{TOOL$\_$3}). Looking at (\ref{eq:26-01-09}), one can notice that the dependence of the integrand on $u_4$ is polynomial, so that a vector field whose divergence corresponds to that integrand can be easily found. Introducing $\mathcal{P}_5(u_1, u_2, u_3, u_4)$ such that
\begin{equation}
\mathcal{P}_5(u_1, u_2, u_3, u_4) \triangleq \int_0^{u_4} \mathcal{P}_4(u_1, u_2, u_3, u'_4) du'_4.
\end{equation}
It leads to
\begin{equation}
\label{eq:26-01-10}
\mathbf{\nabla} \cdot \Big[R^{(i-1)} \mathcal{P}_5 \hat{u}_4 \Big] = R^{(i-1)} \mathcal{P}_4.
\end{equation}

Substituting (\ref{eq:26-01-10}) into (\ref{eq:26-01-09}) and applying the divergence theorem (\emph{TOOL$\_$3}), it can be proven (cf. Appendix) that no special care is required for the possibly singular zone where $R \rightarrow 0$. This zone can be removed from the domain of integration, leading to
\begin{equation}
\label{eq:26-01-11}
\begin{split}
\mathcal{A}_i = \sum_p (\hat{n}_p^4 \cdot \hat{u}_4) \dashint_{\Omega_p^4} &R^{(i-1)}(u_1, u_2, u_3)  \\
&\times \mathcal{P}_5(u_1, u_2, u_3, u_4) dV_3^4,
\end{split}
\end{equation}
with $\Omega_p^6$ the $p^{th}$ 3D polyhedron that is part of the boundary of $\Omega_v^6$, the original volume of integration (cf. Section \ref{sec:6Dgeom}), $\Omega_p^4$ the projection of $\Omega_p^6$ into the subspace $u_6 = u_5 = 0$, $\hat{n}_p^4$ the outer normal to $\hat{n}_p^4$ and $\dashint$ denoting the integration in the Cauchy principal value sens. $\hat{n}_p^4$ can be computed by combining $\hat{n}_p^6$ with $\hat{n}_{v,2}^5$ and $\hat{n}_{v,1}^5$ to remove its $u_6$ and $u_5$ components: 
\begin{equation}
\begin{split}
\mathbf{n}_p^4 = 
	\hat{n}_p ^6
	&- \dfrac{\hat{n}_p^6 \cdot \hat{u}_6}{\hat{n}_{v,2}^5 \cdot \hat{u}_6}
		 \hat{n}_{v,2}^5  \\
	&- \bigg(\dfrac{\hat{n}_p^6 \cdot \hat{u}_5}{\hat{n}_{v,1}^5 \cdot \hat{u}_5}
			-\dfrac{\hat{n}_p^6 \cdot \hat{u}_6}{\hat{n}_{v,1}^5 \cdot \hat{u}_5}
			 \dfrac{\hat{n}_{v,2}^5 \cdot \hat{u}_5}{\hat{n}_{v,2}^5 \cdot \hat{u}_6} \bigg)
		 \hat{n}_{v,1}^5
		 \end{split}
\end{equation}
and $\hat{n}_p^4 = \mathbf{n}_p^4/ |\mathbf{n}_p^4|$.

Using a development similar to the one in the Appendix, it can be shown that the value of the integrals of (\ref{eq:26-01-11}) are bounded, so that contribution of the terms for which $\hat{n}_p^4 \cdot \hat{u}_4 \rightarrow 0$ vanishes.
Removing the vanishing terms from the summation, (\ref{eq:26-01-11}) becomes
\begin{equation}
\label{eq:26-01-12}
\begin{split}
\mathcal{A}_i = \sum_{(p|\hat{n}_p^4 \cdot \hat{u}_4 \neq 0)} (\hat{n}_p^4 \cdot \hat{u}_4) &\dashint_{\Omega_p^4} R^{(i-1)}(u_1, u_2, u_3)  \\
&\times \mathcal{P}_5(u_1, u_2, u_3, u_4)~ dV_3^4.
\end{split}
\end{equation}

Now that we are dealing with a 3D volume within a 4D space, we can proceed as we did in Sections \ref{sec:6Dto5D} and \ref{sec:5Dto4D}. 

First, the $u_4$ dependence of the integrand of (\ref{eq:26-01-12}) is eliminated (\emph{TOOL$\_$1}) using the fact that, within the volume of integration,
\begin{equation}
u_4 = \dfrac{d_p^4 - \sum_{\alpha=1}^{3} (\hat{n}_p^4 \cdot \hat{u}_\alpha) u_\alpha}{\hat{n}_p^4 \cdot \hat{u}_4},
\end{equation}
After, we project the 3D volumes of integration $\Omega_p^4$ into the $u_4 = 0$ subspace (\emph{TOOL$\_$2}), leads to
\begin{equation}
\label{eq:27-01-01}
\begin{split}
\mathcal{A}_i = \sum_{(p|\hat{n}_p^4 \cdot \hat{u}_4 \neq 0)} \dfrac{(\hat{n}_p^4 \cdot \hat{u}_4)}{|\hat{n}_p^4 \cdot \hat{u}_4|} \dashint_{\Omega_p^3} &R^{(i-1)}(u_1, u_2, u_3)  \\
&\times \mathcal{P}_6(u_1, u_2, u_3) ~ dV_3^3.
\end{split}
\end{equation}

Equation (\ref{eq:27-01-01}) corresponds to a ``classical" volume integral of the $R^{(i-1)}$ function multiplied by a given polynomial. The analytical treatment of that kind of integrals will be tackled in Section \ref{sec:3Dint}. 

\subsection{Parallel BF and TF}
\label{sec:parallel}
In the case where the BF is parallel to the TF,  Equation  (\ref{eq:26-01-07}) in Section \ref{sec:5Dto4D} is no longer valid due to the fact that $\hat{n}_{v,1}^5\cdot  \hat{u}_5 = 0$. Indeed, in the parallel case, $\hat{n}_{v,1}^5 = (0,0, \pm1,0,0,0)$, so that the $u_5$ coordinate is independent from the other coordinates within the volume of integration, in (\ref{eq:26-01-06}).

However, for the same reason, the $\hat{u}_5$ direction is entirely contained within the space spanned by this 4D volume of integration. It will be useful to apply the divergence theorem (\emph{TOOL$\_$3}). Introducing $\mathcal{P}_4^\parallel$ such that 
\begin{equation}
\mathcal{P}_4^\parallel \triangleq \int_0^{u_5} \mathcal{P}_3 ~ du'_5
\end{equation}
leads to
\begin{equation}
\label{eq:16-08-01}
\mathbf{\nabla}\cdot \Big[ R^{(i-1)} \mathcal{P}_4^\parallel \hat{u}_5 \Big] = R^{(i-1)} \mathcal{P}_3.
\end{equation}

Substituting (\ref{eq:16-08-01}) into (\ref{eq:26-01-06}) and using the divergence theorem gives
\begin{equation}
\label{eq:27-01-02}
\begin{split}
\mathcal{A}_i = \sum_{(p|\hat{n}_p^5\cdot \hat{u}_5 \neq 0)} (\hat{n}_p^5 \cdot  &\hat{u}_5) \dashint_{\Omega_p^5} R^{(i-1)}(u_1, u_2, u_3)  \\
&\times \mathcal{P}_4^\parallel(u_1, u_2, u_3, u_4, u_5) ~ dV_3^5,
\end{split}
\end{equation}
with $\hat{n}_p^5$ the outer normal of the 3D volume $p$ within the 5D space. No special care is required for the possibly singular zone where $R \rightarrow 0$ (cf. Appendix). Note that the vanishing terms corresponding to $\hat{n}_p^5\cdot \hat{u}_5 = 0$ have been removed from the summation (cf. Section \ref{sec:4Dto3D}).

Since $\hat{n}_p^5\cdot \hat{u}_5 \neq 0$ for all the remaining terms, $u_5$ is no longer linearly independent from the other coordinates (\emph{TOOL$\_$1}):
\begin{equation}
\label{eq:27-01-03}
u_5 = \dfrac{d_p^5 - \sum_{\alpha=1}^{4} \big(\hat{n}_p^5 \cdot \hat{u}_\alpha\big) u_\alpha}{\hat{n}_p^5 \cdot \hat{u}_5}.
\end{equation}
The substitution of (\ref{eq:27-01-03}) into (\ref{eq:27-01-02}), followed by the orthogonal projection of the 3D volumes of integration into the 4D subspace $u_5=0$ (\emph{TOOL$\_$2}) leads to
\begin{equation}
\label{eq:16-08-02}
\begin{split}
\mathcal{A}_i = \sum_{(p|\hat{n}_p^5\cdot \hat{u}_5 \neq 0)} \dfrac{(\hat{n}_p^5 \cdot \hat{u}_5)}{|\hat{n}_p^5 \cdot \hat{u}_5|} & \dashint_{\Omega_p^4} R^{(i-1)}(u_1, u_2, u_3)  \\
&\times \mathcal{P}_5^\parallel(u_1, u_2, u_3, u_4) dV_3^4,
\end{split}
\end{equation}

We can iterate to transform (\ref{eq:16-08-02}) into a sum of 2D surface integrals taking place in the classical 3D space:
\begin{equation}
\label{eq:27-01-04}
\begin{split}
\mathcal{A}_i = &\sum_{(p|\hat{n}_p^5\cdot \hat{u}_5 \neq 0)} \dfrac{(\hat{n}_p^5 \cdot \hat{u}_5)}{|\hat{n}_p^5 \cdot \hat{u}_5|} 
\sum_{(q|\hat{n}_q^4\cdot \hat{u}_4 \neq 0)}^{\Omega_q^4 \in \partial(\Omega_p^4)} \dfrac{(\hat{n}_q^4 \cdot \hat{u}_4)}{|\hat{n}_q^4 \cdot \hat{u}_4|}   \\
 & \times \dashint_{\Omega_q^3} R^{(i-1)}(u_1, u_2, u_3)
\mathcal{P}_6^\parallel(u_1, u_2, u_3) dS_2^3,
\end{split}
\end{equation}
with $\hat{n}_q^4$ the outward unit vector perpendicular to the surface $\Omega_q^4$ that is part of the boundary of the volume $\Omega_p^4$.

(\ref{eq:27-01-04}) is dealing with the surface integration of the ``classical" $R$ function multiplied by a polynomial function of the coordinates, which will be solved in Section \ref{sec:3Dint}.

\section{Estimating the 3D integral}
\label{sec:3Dint}
\begin{figure}
\center
\begin{tabular}{cc}
\includegraphics[width = 4cm]{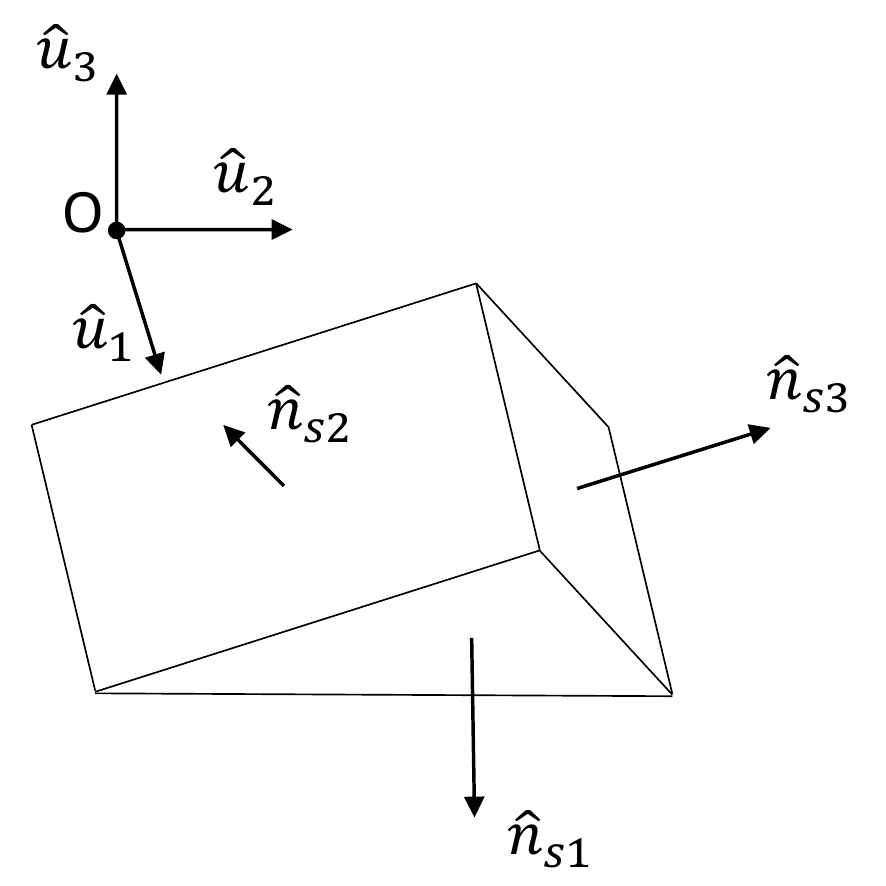}  &
\includegraphics[width = 4cm]{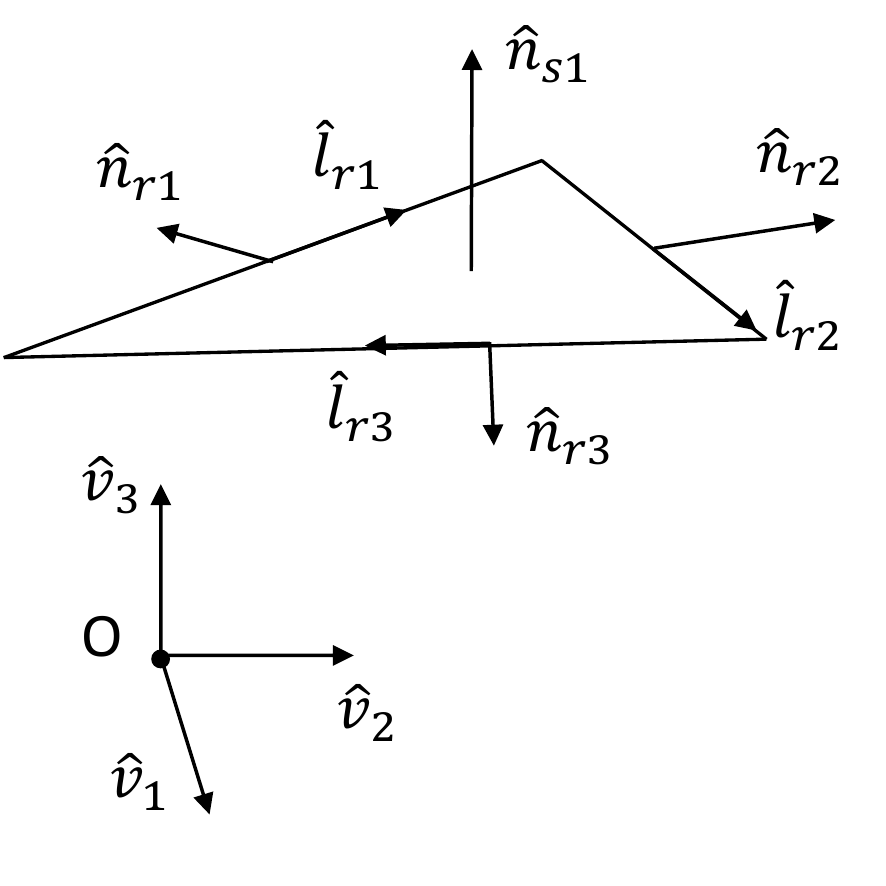} \\
(a) &
(b) 
\end{tabular}\\
\includegraphics[width = 5cm]{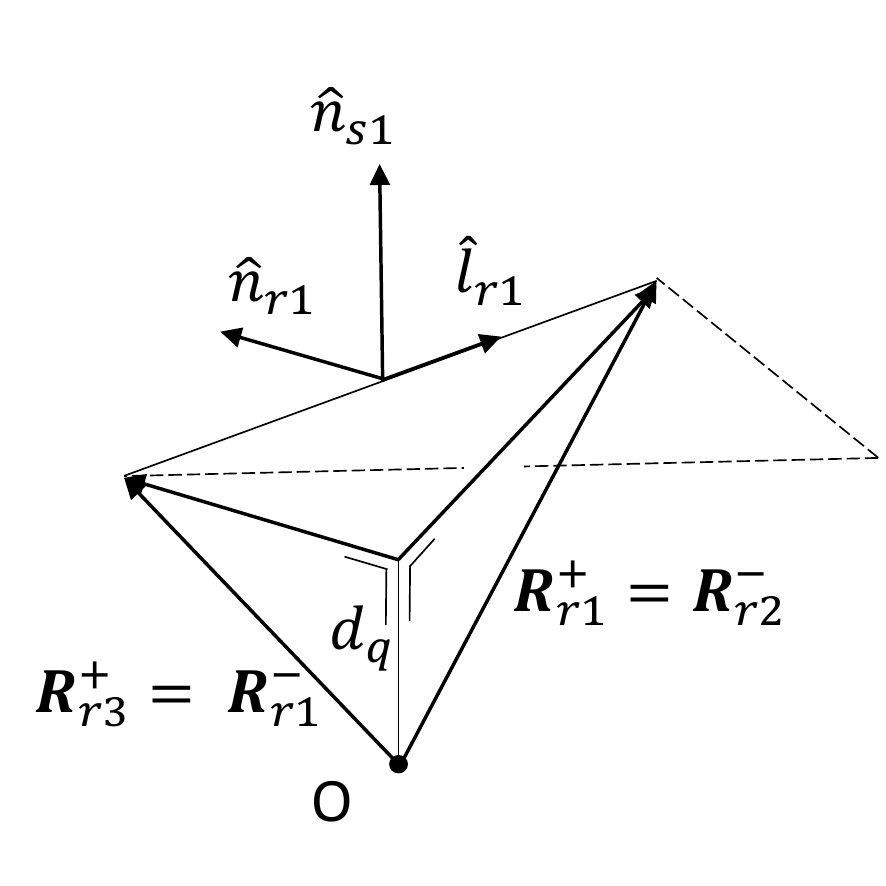} \\
(c)
\caption{Geometry of the (a) 3D, (b) 2D and (c) 1D domains of integration and illustration of the associated quantities.}
\label{fig:struct}
\end{figure}

In this section, we solve the 3D and 2D integrals of Equations (\ref{eq:27-01-01}) and (\ref{eq:27-01-04}). These types of integrals have already been solved by the authors of \cite{Jarvenpaa2003}. The method is summarized here for self-completeness. The geometries described throughout the development are illustrated in Figure \ref{fig:struct}.

We start from (\ref{eq:27-01-01}). The polynomial contribution of some coordinate $u_\alpha$ in $\mathcal{P}_6$ can be isolated by rewriting
\begin{equation}
\label{eq:01-09-01}
\mathcal{P}_6(u_1, u_2, u_3) = \sum_s P_{s}(u_\beta, u_\gamma) u_\alpha^s, ~~~~ \alpha \neq \beta \neq \gamma \neq \alpha.
\end{equation}

Each term of (\ref{eq:01-09-01}) is treated separately. To reduce the dimensionality of the integrals of (\ref{eq:27-01-01}), the following identities are used:

\begin{subequations}
\label{eq:27-01-07}
\underline{If $s>0$:}
\begin{align}
\mathcal{P}_{s}(u_\beta, u_\gamma) u_\alpha^s R^{(\zeta-1)} &= \dfrac{1}{\zeta+1} \nonumber \\
\times \bigg[\mathbf{\nabla} \cdot \Big(\mathcal{P}_{s}(u_\beta, & u_\gamma) u_\alpha^{(s-1)} R^{(\zeta+1)} \hat{u}_\alpha \Big) \label{eq:08-02-04} \\
- (s-1) & \mathcal{P}_{s}(u_\beta, u_\gamma) u_\alpha^{(s-2)} R^{(\zeta+1)} \bigg], \nonumber
\end{align}
\underline{If $s=0$ and $P_s(u_\beta, u_\gamma) = cstt$: }
\begin{align}
R^{(\zeta-1)} &= \dfrac{1}{(\zeta+2)}\mathbf{\nabla} \cdot \Big(R^{\zeta} \hat{R} \Big),
\end{align}
\underline{Otherwise:}
\begin{align}
\mathcal{P}_{s}(u_\beta, u_\gamma) R^{(\zeta-1)} &= \sum_t \mathcal{P}_{t}(u_\gamma) u_\beta^t R^{(\zeta-1)} \label{eq:27-01-08}.
\end{align}
\end{subequations}

Substituting these identities recursively into (\ref{eq:27-01-01}) and applying the divergence theorem, the volume integral can be expressed as a sum of surface integrals. Note that the $R (\hat{R} \cdot \hat{n}_q)$ and $\hat{u}_\alpha \cdot \hat{n}_q$ factors are constant on surface $q$ and can therefore be removed from the integral.

Using the development in the Appendix, it can be shown that no special treatment is required for the possibly singular zone where $R \rightarrow 0$.

We now have to deal with surface integrals of the type
\begin{equation}
\label{eq:08-02-05}
\dashint \mathcal{P}_s(u_1, u_2, u_3) R^{(\zeta-1)} dS, ~~~~ \zeta \geq 0.
\end{equation}
First of all, the $\{\hat{u}_1, \hat{u}_2, \hat{u}_3\}$ orthonormal right-handed basis is rotated to match the $\{\hat{v}_1, \hat{v}_2, \hat{v}_3\}$ orthonormal right-handed basis, with $\hat{v}_1$ and $\hat{v}_2$ parallel to the surface. In this new basis, the $v_3$ coordinate is constant along the surface of integration, so that 
\begin{equation}
\label{eq:27-01-09}
\dashint \mathcal{P}_s(u_1, u_2, u_3) R^{(\zeta-1)} dS = \dashint \mathcal{P}_s^v(v_1, v_2) R^{(\zeta-1)} dS.
\end{equation}

To reduce the 2D integral of (\ref{eq:27-01-09}) into a sum of 1D integrals, the different terms treated separately using (\ref{eq:27-01-08}) and the following identities are recursively used:

\begin{subequations}
\underline{If $s>0$:}
\begin{align}
\mathcal{P}_{s}^v(v_\beta) v_\alpha^s R^{(\zeta-1)} = \dfrac{1}{\zeta+1}& \nonumber \\
\times \bigg[\mathbf{\nabla} \cdot \Big(\mathcal{P}_{s}^v(& v_\beta) v_\alpha^{(s-1)} R^{(\zeta+1)} \hat{v}_\alpha \Big)  \\
- (s-1) & \mathcal{P}_{s}^v(v_\beta) v_\alpha^{(s-2)} R^{(\zeta+1)} \bigg], \nonumber
\end{align}
\underline{If $s=0$ and $P_s(v_\beta) = cstt$:}
\begin{align}
R^{(\zeta-1)} &= \dfrac{1}{\zeta+1} \mathbf{\nabla} \cdot \bigg( \dfrac{R^{(\zeta+1)}}{P_q^2} \mathbf{P}_q \bigg), \label{eq:06-02-01}
\end{align}
\underline{Otherwise:}
\begin{align}
\mathcal{P}_{s}^{v}(v_\beta) R^{(\zeta-1)} &= \sum_w K_w v_\beta^w R^{(\zeta-1)},
\end{align}
\end{subequations}
with $K_w$ a constant, $\mathbf{P}_q = (v_1 \hat{v}_1 +  v_2 \hat{v}_2)$, which corresponds to the projection of $\mathbf{R}$ into the plane of the surface $q$, and $P_q = |\mathbf{P}_q|$. 1D integrals are obtained after application of the divergence theorem. In those integrals, denoting by $\hat{n}_r$ the outward unit normal to the edge $r$, the factors $(\mathbf{P}_q\cdot \hat{n}_r)$ and $\hat{v}_\alpha \cdot \hat{n}_r$ are constant along the edge, so that they can be moved outside the integrals.

As explained in \cite{Wilton84, Caorsi1993sept}, a special care is required if the surface contains one point for which $P_q = 0$. Indeed, in that case, the vector field of (\ref{eq:06-02-01}) becomes singular. Following \cite{Caorsi1993sept}, separating the integral into two parts, it can be shown that
\begin{align}
\label{eq:07-02-01}
&\int_{\Omega_q^3} \mathbf{\nabla}\cdot  \bigg( \dfrac{R^{(\zeta+1)}}{P_q^2} \mathbf{P}_q \bigg) dS_2^3 
= \nonumber\\
&\sum_{r | \Omega_r^3 \in \partial \Omega_q^3} \Bigg[ (\mathbf{P}_q \cdot \hat{n}_r) \int_{l_r}  \dfrac{R^{(\zeta+1)}}{P_q^2} dl_1^3 \\
&-
d_q^{(\zeta+1)}
\bigg(\tan^{-1} \bigg( \dfrac{\hat{l}_r\cdot \mathbf{R}_r^+}{\hat{n}_r \cdot \mathbf{R}_r^+} \bigg) - \tan^{-1} \bigg(\dfrac{\hat{l}_r\cdot \mathbf{R}_r^-}{\hat{n}_r \cdot \mathbf{R}_r^-}\bigg) \bigg) \Bigg], \nonumber
\end{align}
with $d_q = |\hat{n}_q \cdot \mathbf{R}| \text{ for } \mathbf{R}\in \Omega_q^3$ the distance between the surface and the origin, $\Omega_r^3$ the projection of edge $r$ into the 3D space, $\mathbf{R}_r^-$ and $\mathbf{R}_r^+$ the position of the beginning and end of $\Omega_r^3$, respectively, and $\hat{l}_r = (\mathbf{R}_r^+ - \mathbf{R}_r^-)/|\mathbf{R}_r^+ - \mathbf{R}_r^-|$. The beginning of one edge is chosen to always correspond to the end of another one.  The second term of the RHS of (\ref{eq:07-02-01}) accounts for the singularity if the point $P_q = 0$ is on the surface, and otherwise vanishes.
For more details about the development that leads to (\ref{eq:07-02-01}), the reader is referred to Equation (5) of \cite{Wilton84} or Equations (39) and (40) of \cite{Jarvenpaa2003}. In this work, contrarily to \cite{Wilton84, Jarvenpaa2003}, Equation (\ref{eq:07-02-01}) is used for both the EFIE and MFIE terms when the whole 4D integral is evaluated analytically.

Last comes the 1D integration. Following the same guidelines as for the 3D and 2D integrations, we use the $\{\hat{l}_r, \hat{n}_r, \hat{n}_q \}$ coordinate system, so that the 1D integrals that need to be solved all correspond to
\begin{equation}
\dashint l^n R^{(\zeta+1)} dl
\end{equation}
or 
\begin{equation}
(\mathbf{P}_q \cdot \hat{n}_r) \dashint \dfrac{R^{(\zeta+1)}}{P_q^2} dl,
\end{equation}
with $l = \mathbf{R} \cdot \hat{l}_r$. If $\mathbf{P}_q \cdot \hat{n}_r = 0$ on an edge, the first and second terms of the right-hand side of (\ref{eq:07-02-01}) mutually cancel for that edge, so that this case can be treated by omitting the whole contribution of the edge to (\ref{eq:07-02-01}).

We can solve these integrals recursively using the following identities:
\begin{subequations}
\begin{align}
 &\int \dfrac{R^{(\zeta+1)}}{P_q^2} dl= d_q^2 \int \dfrac{R^{(\zeta-1)}}{P_q^2} dl + \int R^{(\zeta-1)} dl, \label{eq:07-02-02} \\
&\int R^{(\zeta+1)} dl = (d_q^2+d_r^2) \int R^{(\zeta-1)} dl \label{eq:07-02-03}  \\
&~~~~~~~~~~~~~~~~~~~~~~~~~~~~~~~~ + \int l^2 R^{(\zeta-1)} dl, \nonumber\\
&\int l^n R^{(\zeta+1)} dl = \Bigg[\dfrac{l^{(n-1)} R^{(\zeta+3)}}{\zeta +3} \Bigg]_{\mathbf{R}^-}^{\mathbf{R}^+} \\
&~~~~~~~~~~~~~~~~~~~~~~ - \dfrac{n-1}{\zeta+3} \int l^{(n-2)} R^{(\zeta + 3)} dl, \nonumber \\
& \int R^{-2} dl = \Bigg[\dfrac{1}{\sqrt{d_q^2+d_r^2}} \tan^{-1}\bigg(\dfrac{l}{\sqrt{d_q^2+d_r^2}}\bigg) \Bigg]_{\mathbf{R}^-}^{\mathbf{R}^+}, \\
&\int R^{-1} dl  = \bigg[\ln(R+l) \bigg]_{\mathbf{R}^-}^{\mathbf{R}^+}, \\
&\int \dfrac{R^{-1}}{P_q^2} dl = \bigg[\dfrac{1}{d_q d_r} \tan^{-1}\bigg(\dfrac{d_q l}{d_r R} \bigg) \bigg]_{\mathbf{R}^-}^{\mathbf{R}^+}, \label{eq:08-02-02}\\
&\int \dfrac{1}{P_q^2} dl = \bigg[\dfrac{1}{d_r} \tan^{-1}\bigg(\dfrac{l}{d_r} \bigg) \bigg]_{\mathbf{R}^-}^{\mathbf{R}^+},
\end{align}
\end{subequations}
with $[f(\mathbf{R})]_{\mathbf{R}_1}^{\mathbf{R}_2} = f(\mathbf{R}_2)-f(\mathbf{R}_1)$ and $d_r = |\mathbf{R} \cdot \hat{n}_r|$. (\ref{eq:07-02-02}) is obtained noticing that $R^2= P_q^2+ d_q ^2$ and that $d_q$ is constant along the edge. (\ref{eq:07-02-03}) is obtained noticing that $d_r$ is also constant along the edge. The case $d_q \rightarrow 0$ in (\ref{eq:08-02-02}) can be straightforwardly treated considering that $P_q \rightarrow R$ and manipulating (\ref{eq:07-02-03}) to increase the power of $R^{-3}$.

This last step ends our quest for the 4D integral. It is important to notice that among the large number of terms that need to be computed to solve the original 4D integral, there are many repetitions. Therefore, using an appropriate book-keeping, a significant time saving can be obtained.

\section{Solving the $\mathcal{C}_i$ MFIE term}
\label{sec:Ci}
The solution of the $\mathcal{C}_i$ term, which is necessary when setting up the discretized MFIE, is similar to the solution of the $\mathcal{A}_{i-1}$ term. The only difference lies in the treatment of the singular case $R \rightarrow 0$, which is described in this Section.

To treat the singular case for the $\mathcal{A}_i$ term (cf. Appendix), we first separated the domain of integration into two parts: the singular and the non-singular zones. Since (\ref{eq:Ci}) correspond to a Cauchy integral, only the non-singular zone needs to be treated; more specifically the additional surface $S_a$ that is created when the singular zone is removed from the volume of integration (cf. Figure \ref{fig:div_geom}). 

First, since the normal to $S_a$ is orthogonal to the $u_4$ and $u_5$ directions, no additional treatment is required w.r.t. Section \ref{sec:6Dto3D}. Moreover, since it is well known that (\ref{eq:Ci}) vanishes for BF and TF lying in the same plane, contrary to the EFIE case, the case of touching parallel BF and TF does not need to be treated, leading to an integral of the type:
\begin{equation}
\label{eq:08-02-03}
\dashint R^{(i-3)} \big(\mathbf{R} \cdot \mathbf{P}_2(u_1, u_2, u_3) \big) dV_3^3.
\end{equation}

The factor $\mathbf{R} \cdot \mathbf{P}_2$ corresponds to a polynomial that vanishes for $R \rightarrow 0$, i.e. a polynomial whose minimum degree is 1.
Therefore, in order to apply the divergence theorem to (\ref{eq:08-02-03}), the identity (\ref{eq:08-02-04}) must be used at least once. After using it, it can be seen that the power of the distance term is increased: $R^{(i-3)}$ becomes $R^{(i-1)}$. The resulting 2D integrals thus match the generic shape of Equation (\ref{eq:08-02-05}), which has already been treated in Section \ref{sec:3Dint}. Moreover, as explained in the Appendix, the contribution of $S_a$ to the 3D volume integral vanishes for a vanishing singular zone.

\section{Numerical validation}
\label{sec:numerical_validation}
\begin{figure}[h!]
\center
\includegraphics[width = 6cm]{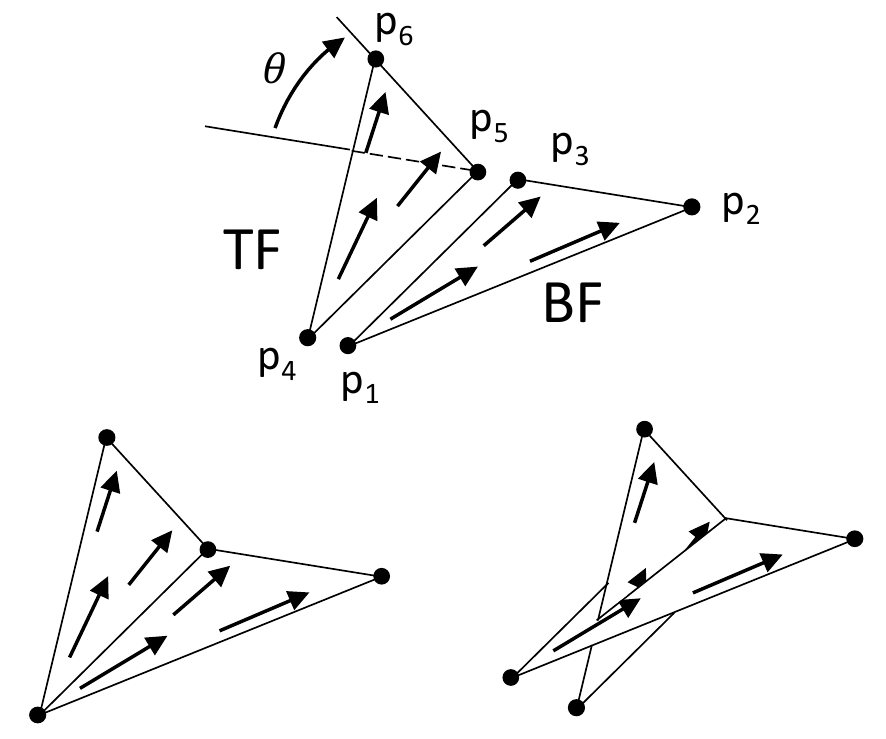} \\
(a)\\
\begin{tabular}{cc}
\hspace{-0.65cm}
\includegraphics[width = 4cm]{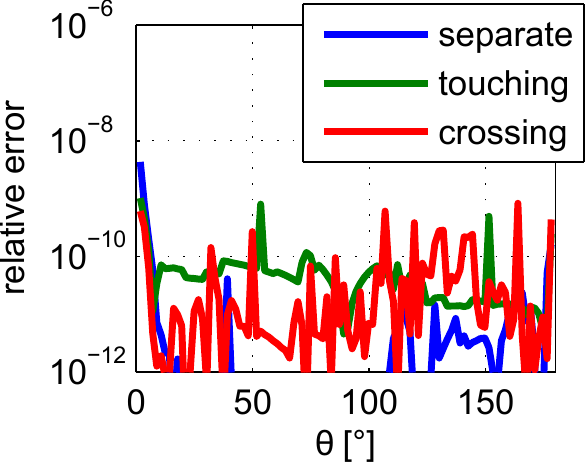}
&
\includegraphics[width = 4cm]{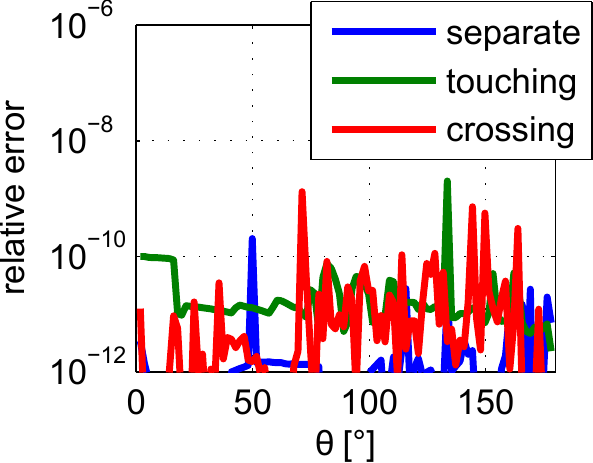} \\
(b) & \hspace{0.2cm} (c)
\end{tabular}
\hspace{-0.65cm}
\includegraphics[width = 4cm]{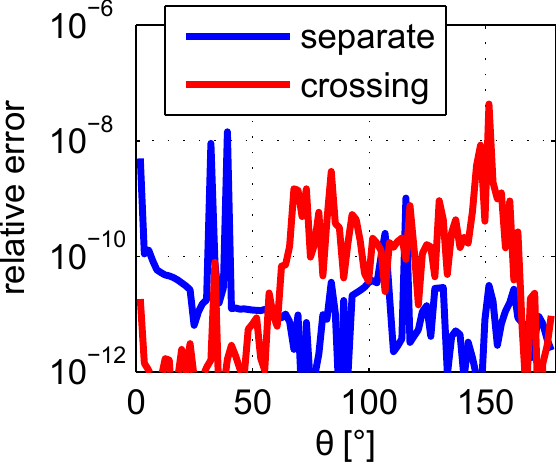} \\
\hspace{0.6cm}(d)
\caption{Comparison between analytical results and numerical results for the (b) $A_0$, (c) $B_0$ and (d) $C_0$ terms for the three geometries illustrated in (a).}
\label{fig:accuracy}
\end{figure}

In order to validate the method, we first computed the $A_0$, $B_0$ and $C_0$ terms of (\ref{eq:04-04-01}). The reference results were obtained using the methods of \cite{Wilton84, Graglia1993, Jarvenpaa2003}, which consist of a 2D analytical integration of the singular term over the BF and the 2D numerical integration of the result over the TF. The numerical integration was performed using the \texttt{integral2} function of MATLAB$^\text{\textregistered}$ \cite{integral2, integral2bis}. We computed these terms for half-RWG \cite{RWG} BF and TF that are close by, sharing a common edge or even interleaved, and for various relative orientations. The geometry considered can be seen in Figure \ref{fig:accuracy}. The position of the vertices is given by
\begin{equation}
\begin{array}{lcl}
\mathbf{p}_1 = \big(0, 0, 0\big),
&  &
\mathbf{p}_2 = \big(1, 1, 0\big), \\
\mathbf{p}_3 = \big(1, 0, 0\big), 
&  &
\mathbf{p}_4 = \mathbf{p}_1+\mathbf{t},\\
\mathbf{p}_5 = \mathbf{p}_3 + \mathbf{t},
&  &
\mathbf{p}_6 = \big(-\cos\theta, 1, \sin\theta\big) + \mathbf{t},
\end{array}
\end{equation}
with $\mathbf{t} = (-0.1, 0, 0)$ when the BF and TF are nearby, $\mathbf{t} = (0, 0, 0)$ when they are touching and $\mathbf{t} = (0.3, 0, -0.2)$ when they are interleaved. The geometry and the results for the three cases can be seen in Fig. \ref{fig:accuracy}. The error in $A_0$, $B_0$ and $C_0$ can be seen in Fig. \ref{fig:accuracy}(b), (c) and (d), respectively. Note that if $\mathbf{t} = (0, 0, 0)$ (common edge), $C_0 = 0$. The results are therefore not displayed for that case in Figure \ref{fig:accuracy}(d).

The error has been normalized w.r.t. the maximum value obtained for each term.
Whatever the geometry considered, the relative error remains below $10^{-7}$ and is generally smaller than $10^{-10}$. It corresponds to the maximum accuracy we could reach for the numerical solution used as a reference. 

\begin{figure}[h!]
\center
\includegraphics[width = 5cm]{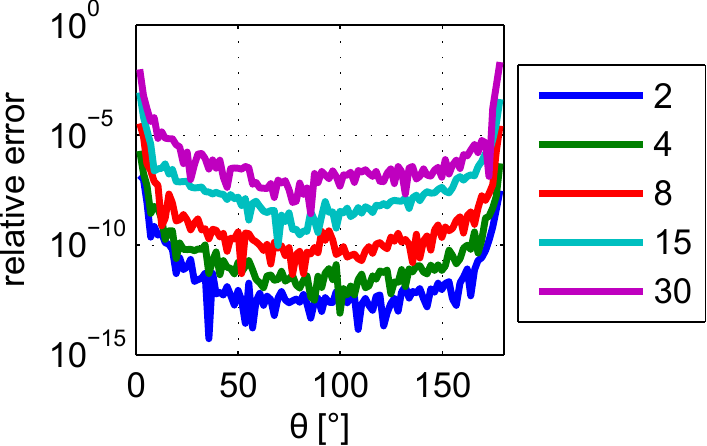} \\
\hspace{-0.5cm}(a)\\
\includegraphics[width = 8.5cm]{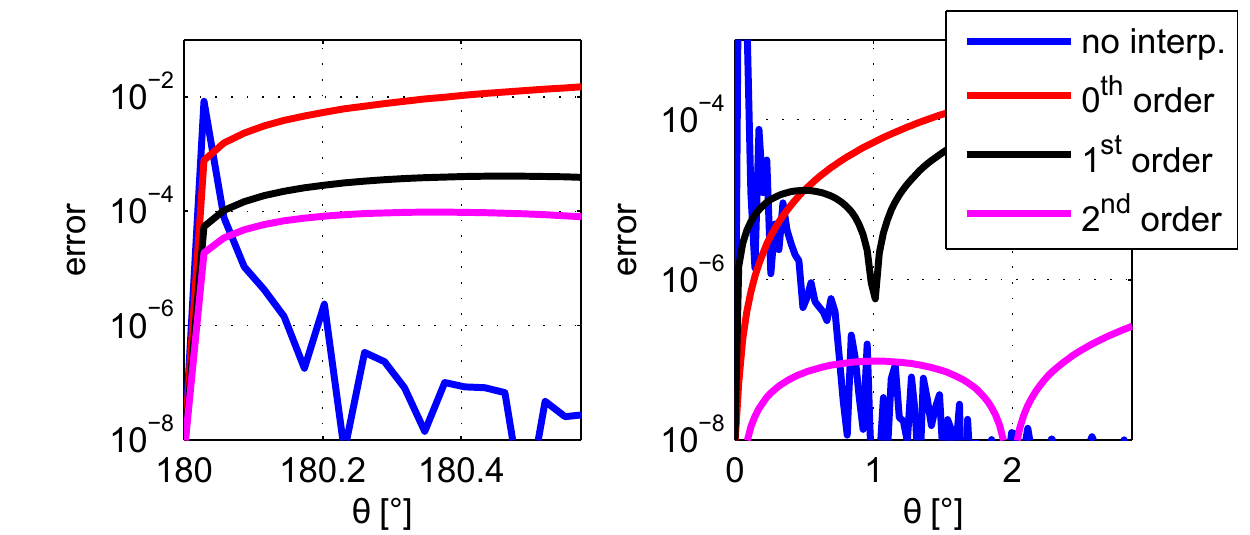} \\
\hspace{0.5cm} (b)  \hspace{3.1cm} (c)
\caption{Round-off error in the $A_0$ term for (a) an increasing distance between the BF and TF and (b,c) for nearly parallel BF and TF sharing a common edge. In the latter graphs, the error made using polynomial interpolation techniques is also drawn.}
\label{fig:limits}
\end{figure}

When the distance between the BF and TF becomes large, some of the analytical integration steps become numerically ill-conditioned, i.e. they involve differences between large values of same order. When the BF and TF are nearly parallel, the non-parallel case becomes ill-conditioned as well. Therefore, we studied both limits, the results for the $A_0$ term being displayed in Fig. \ref{fig:limits}. Note that a similar behaviour was observed for the $A_i$ and $C_i$ terms ($i\geq0$) while a much better accuracy was observed for the $B_i$ terms. The latter observation is due to the degree of the polynomial in (\ref{eq:25-01-03}), which is 2 for the $A_i$ and $C_i$ terms, and 0 for the $B_i$ term in the case of RWG basis functions. 

First, we study the numerical error for increasing distances between the BF and TF. We studied the same configuration as in the previous example (Fig. \ref{fig:accuracy}(a)) with $\mathbf{t} = (-t, 0, 0)$ and $t$ = 2, 4, 8, 15 and 30. As shown in Fig. \ref{fig:limits}(a), the accuracy rapidly drops if the distance between the BF and TF becomes large ($t$ larger than 5 to 10 times the typical size of an edge of the mesh). This is not expected to pose a problem in MoM applications since at such distances classical Gaussian quadratures perform very well (see e.g. \cite{Wang2003}). For this reason, the analytical method should be used only for close-range singular interactions, smaller than about 5 times the typical length scale of the mesh.

Second, we studied the evolution of the error for small angles between the BF and TF. To do so, we considered again a geometry similar to the first example (Fig. \ref{fig:accuracy}(a)) with $\mathbf{t} = (0,0,0)$. The error for nearly-parallel BF and TF is displayed in blue in Fig. \ref{fig:limits}(b-c). Note that $\theta = 180^\circ$ corresponds to the TF that has been folded back into the BF (identical BF and TF). The other curves correspond to polynomial interpolation of order $n$ of the value obtained for $\theta = 0^\circ, 1^\circ,... n^\circ$ and  $\theta = 180^\circ, 181^\circ,... 180+n^\circ$ (cf. end of Section \ref{sec:5Dto4D}). An accuracy better than $10^{-4}$ can be easily achieved using a second order interpolation. 

\begin{figure}[h!]
\center
\includegraphics[width = 8cm]{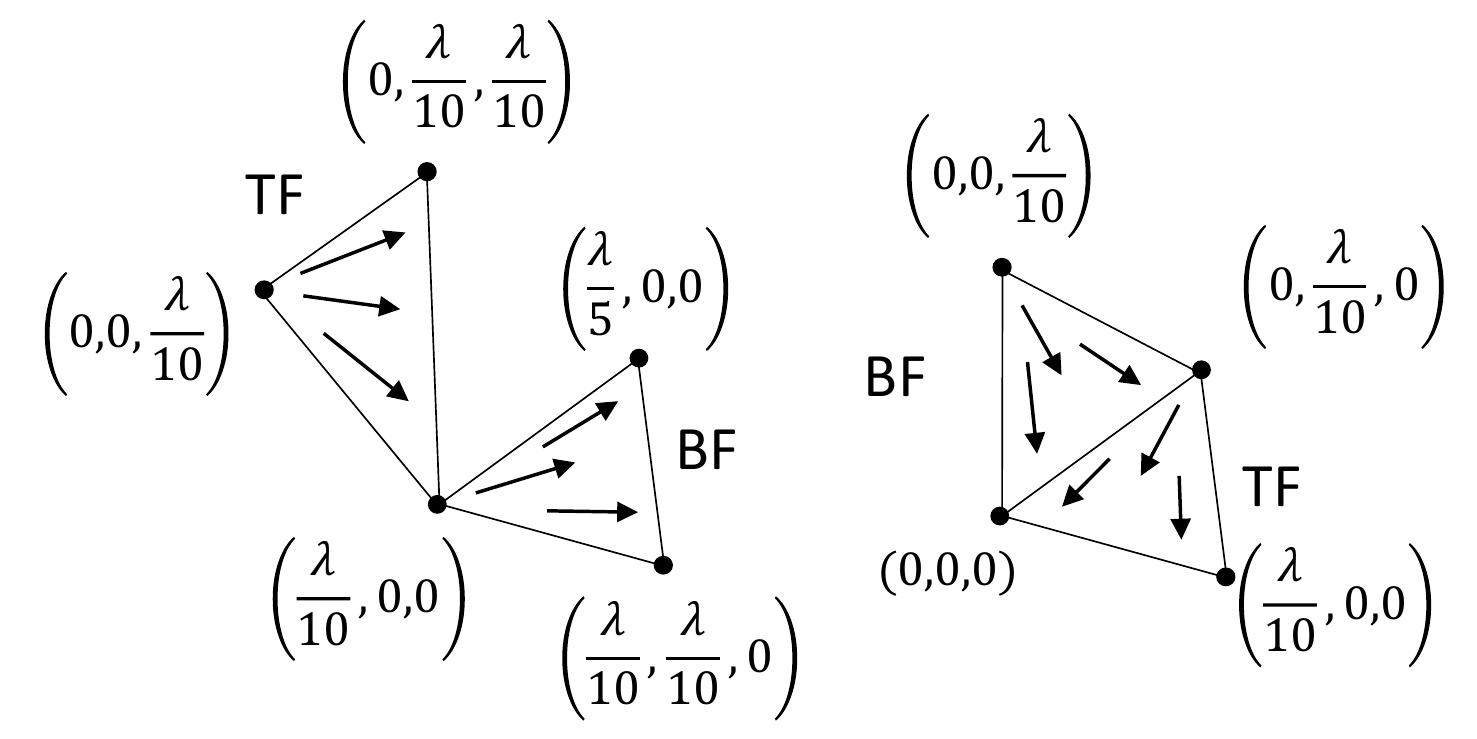} \\
(a) \\
\vspace{0.3cm}
\begin{tabular}{cc}
\includegraphics[width = 4cm]{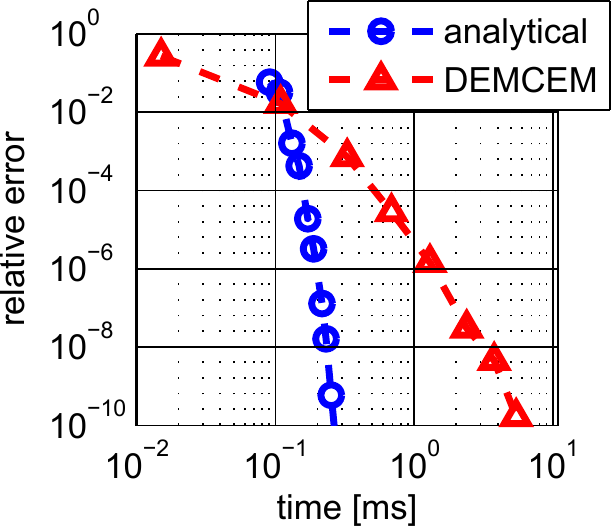} \hspace{-0.3cm}
&
\includegraphics[width = 4cm]{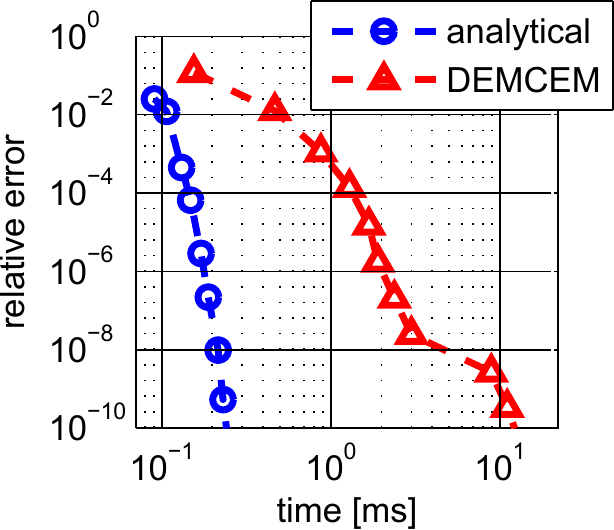} \\
\hspace{0.6cm}(b) & \hspace{0.6cm}(c) \\
\includegraphics[width = 4cm]{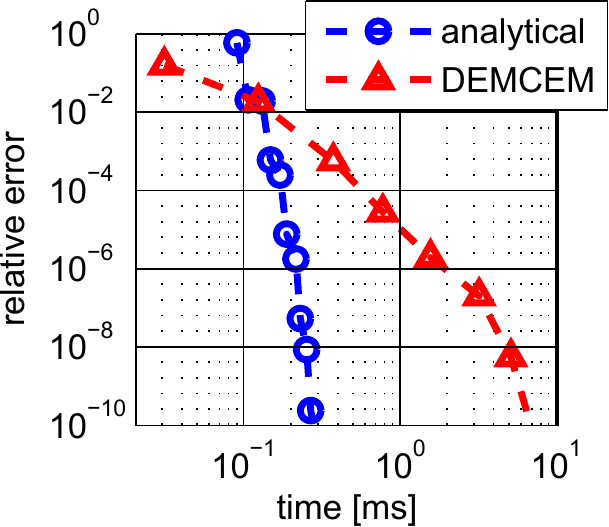} \hspace{-0.3cm}
&
\includegraphics[width = 4cm]{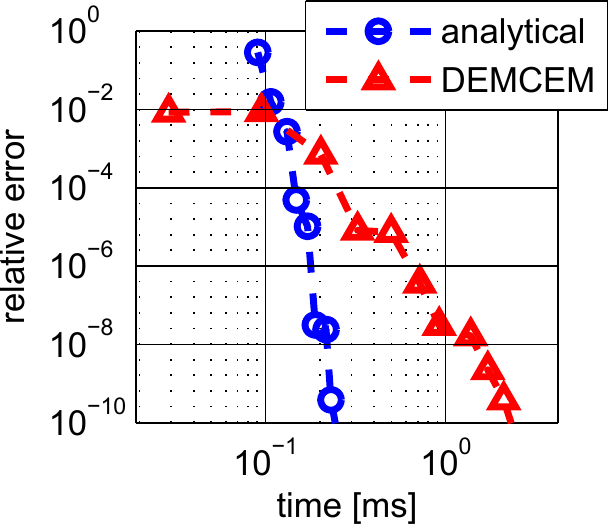} \\
\hspace{0.6cm}(d) & \hspace{0.6cm}(e) 
\end{tabular}
\caption{Comparison of the proposed method with the method of \cite{DEMCEM} for the two geometries illustrated in (a). (b) $Z^{EJ}$ and (d) $Z^{EM}$ for BF and TF with a common vertex. (c) $Z^{EJ}$ and $Z^{EM}$ for BF and TF sharing a common edge.}
\label{fig:polim}
\end{figure}

Finally, in order to compare our method with the state of the art for practical geometries, we computed $Z^{EJ}$ and $Z^{EM}$ of (\ref{eq:25-01-01}) using the proposed method (in-house C++ code)
and the ``Direct Evaluation Method" (DEMCEM) of A. Polimeridis \cite{Polimeridis2010march, Polimeridis2011, Polimeridis2008sept, Polimeridis2011june}, whose C++ implementation can be found on the website of the author \cite{DEMCEM}. Both codes were compiled using the \texttt{-O2} optimization flag and run on a single CPU of an Intel i7-3770 chip. The accuracy was checked for varying numbers of points for the 1D numerical integrations in \cite{DEMCEM} and various orders of the Taylor expansion for the proposed method ($N$ in Equation (\ref{eq:04-04-02})). The comparison was lead for BF and TF sharing a common edge or a common vertex, as illustrated in Fig. \ref{fig:polim}(a). 

To obtain a fair comparison, the time required to compute the value of one integral was averaged over $10^4$ integrals (100 BF interacting with 100 TF).
It is worth noting that the results displayed varied slightly from run to run and may change for different numerical implementations of both methods. However, comparisons have been made as fair as possible, considering that both codes are reasonably optimized and have been compiled and executed on the same computer and the results are averaged over a large number of different interactions.

The times provided for the fully analytical method are actually covering the computation of the two integrals of (\ref{eq:04-04-02}), since many factors are reused and therefore both calculations cannot be decoupled efficiently. The results can be seen in Figure \ref{fig:polim}(b-e). The reference solution used to estimate the error corresponds to the results obtained using the code of \cite{DEMCEM} for 15 points of integration. 
The first bullet of the DEMCEM results corresponds to one point of integration for each 1D numerical integral, the second one to two points of integration for each 1D numerical integral, etc. Concerning the proposed method, the first bullet corresponds to $N=0$, the second one to $N=1$, etc. Relative errors of the order of $10^{-10}$ are achieved 10 to 40 times faster than with the code of \cite{DEMCEM}. 

 The convergence of the fully analytical method is not expected to change significantly with respect to the distance between the BF and TF (except for the numerical error that may grow, see Fig. \ref{fig:limits}(a)) since $R_0$, the central point of the Taylor expansion of Equation (\ref{eq:04-04-02}), is changing accordingly \cite{FMIR}. 
 However, it will strongly vary with the density of the mesh. For BF and TF whose characteristic length is $\lambda/10$, the trends shown in Fig. \ref{fig:polim} are typical and an accuracy of $10^{-10}$ is achieved with 9 to 11 terms ($N=8$ to $10$).

\section{Conclusion}
In this paper, we provided a fully analytical solution to the 4D singular integrals that appear in the static kernel of the PMCHWT formulation of the MoM involving both EFIE and MFIE operators.
First, the double 2D integrals over the BF and TF taking place in the 3D classical space are reformulated into a 4D integral over a 4D volume of integration lying inside a 6D space. Using a sum-and-difference change of coordinates followed by alternating use of the divergence theorem and orthogonal projection of the volume of integration into smaller subspaces, the 6D integral can be reformulated as a sum of 3D integrals taking place in the classical 3D space. These integrals have already been solved analytically in the literature \cite{Jarvenpaa2003}, leading to a final closed-form expression.

Using a rapidly converging Taylor expansion of the phase factor that appears in the dynamic kernel \cite{FMIR}, the presented method can be straightforwardly extended to the computation of the 4D integral arising in the dynamic kernel of the MoM.
The method presented is applicable to any kind of flat polygonal BF and TF of any order and can be straightforwardly extended to the volumetric MoM. Moreover, the time required to analytically evaluate the 4D singular integrals proves to be competitive with efficient numerical techniques, such as \cite{DEMCEM}.

This method should be used to compute the close-range interactions between BF and TF, i.e. for BF and TF separated by less than about 5 edge lengths. It is not suited in its present form for distant BF and TF since some of the operations involved in the 4D analytical integration are ill-conditioned. However, for these distant interactions, the integrand is not singular and classical Gaussian quadratures rapidly provide accurate results.

\section*{Acknowledgments}
The authors would like to thank Athanasios Polimeridis for sharing his code through his website.

\begin{appendix}
In this section, we will show that, most of the time, no particular treatment is required for the possibly singular region where $R\rightarrow 0$. This section follows the same reasoning as \cite{Wilton84}, but extends it to higher dimensions.

Since the divergence theorem is only valid for sufficiently smooth functions, singular integrands need a special care. Following a ``divide-and-conquer" strategy, the integration domain is split into two parts, one corresponding to the singular region of vanishing extent and the other one corresponding to the rest of the domain of integration, over which the integrand is regular.

Consider the regular function $f$ and the vector field $\mathbf{g}_i$ such that
\begin{equation}
\label{eq:05-04-02}
f(\mathbf{u}) R^{(i-1)}(\mathbf{u}) = \mathbf{\nabla} \cdot \mathbf{g}_i (\mathbf{u}), ~~~~ \forall \mathbf{u} \in \{\mathbf{u}|R(\mathbf{u})\geq R_0\}.
\end{equation}
with $R$ the distance function, $R_0>0$ and $i\geq 0$. In the following, we will consider the limit $R_0 \rightarrow 0^+$. We want to solve the following integral over a $k$-D domain $\Omega$:
\begin{equation}
\label{eq:05-04-01}
I_S = \int_\Omega f R^{(i-1)} dV_k^k.
\end{equation}

\begin{figure}[h!]
\center
\includegraphics[width = 8cm]{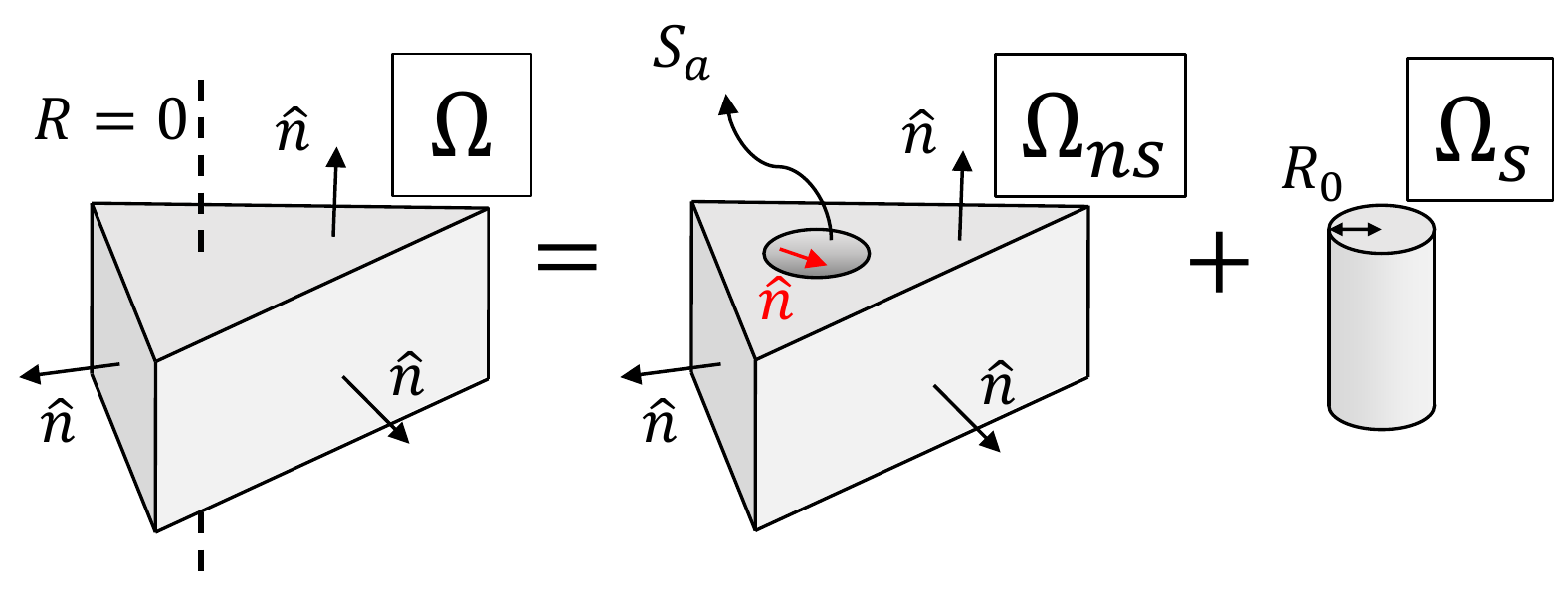}
\caption{Singularity treatment of a 3D volume with the $R$ function varying only along two dimensions.}
\label{fig:div_geom}
\end{figure}

First, we separate the domain of integration $\Omega$ into two parts, $\Omega_{\text{ns}}$ and $\Omega_{\text{s}}$, such that
\begin{subequations}
\begin{align}
\Omega_{\text{s}} &= \Omega \cap \{\mathbf{u}|R(\mathbf{u}) < R_0\}, \\
\Omega_\text{ns} &= \Omega \backslash \Omega_\text{s}.
\end{align}
\end{subequations}
In Figure \ref{fig:div_geom}, an example is provided for a 3D volume with a distance function varying along two out of the three coordinates.
Splitting the integral of (\ref{eq:05-04-01}) and substituting (\ref{eq:05-04-02}) where the equation is valid gives
\begin{equation}
I_S = \int_{\Omega_\text{ns}} \mathbf{\nabla} \cdot \mathbf{g} ~ dV_k^k + \int_{\Omega_\text{s}} f R^{(i-1)} ~ dV_k^k.
\end{equation}
The goal of this section is to provide a sufficient condition to simply neglect the singular zone, i.e. a sufficient condition for the following identity to hold true:
\begin{equation}
\label{eq:05-04-05}
\int_{\Omega_\text{ns}} \mathbf{\nabla} \cdot \mathbf{g} ~ dV_k^k + \int_{\Omega_\text{s}} f R^{(i-1)} ~ dV_k^k =
\dashint_{\partial \Omega} \mathbf{g} \cdot \hat{n} ~ dV_{k-1}^k,
\end{equation} 
with $\partial \Omega$ the surface of $\Omega$ and $\hat{n}$ the outer unit normal of $\partial \Omega$. First, we will derive a sufficient condition for the contribution of the integral over $\Omega_\text{s}$ to be negligible. Then, we will tackle the contribution of the surface $\{\partial \Omega_\text{ns}\} \backslash  \{\partial \Omega\}$ (surface $S_a$ in Figure \ref{fig:div_geom}).

\paragraph{Singular volume} 
Consider the shape of the volume $\Omega_\text{s}$. Along the $\hat{u}_{1\rightarrow3}$ directions, which the $R$ function depends on, it will correspond to a circle (2D) or a sphere (3D) of radius $R_0$. Along the $\hat{u}_{4\rightarrow6}$ directions, the volume corresponds to a line (1D), polygon (2D) or polyhedron (3D) whose length, surface or volume will be denoted by $h$. Figure \ref{fig:div_geom} depicts the case of a 3D volume, with a $R$ function that only varies along two out of the three dimensions spanned by the volume.

We introduce $f_\text{max}$ such that $\forall \mathbf{u} \in \Omega$, we have $|f(\mathbf{u})| < f_\text{max}$. We can then write
\begin{subequations}
\begin{align}
\Bigg|\int_{\Omega_\text{s}} f R^{(i-1)} ~ dV_k^k\Bigg|
& \leq \int_{\Omega_\text{s}} \big|f R^{(i-1)}\big|~dV_k^k \\
& < f_\text{max} \int_{\Omega_\text{s}} R^{(i-1)} ~ dV_k^k. \label{eq:05-04-03}
\end{align}
\end{subequations}

It can be noticed that the integrand of the RHS of (\ref{eq:05-04-03}) is only depending on $R$, so that the integration over dimensions $\hat{u}_{4\rightarrow6}$ can be done in a trivial way:
\begin{equation}
f_\text{max} \int_{\Omega_\text{s}} R^{(i-1)} ~ dV_k^k
=
f_\text{max} h \int_{\Omega^R_\text{s}} R^{(i-1)} ~ dV_N^3,
\end{equation}
with $\Omega^R_\text{s}$ the section of $\Omega_\text{s}$ in the $\hat{u}_{1\rightarrow3}$ space and $N$ the dimensionality of this section ($N=2$ for a circle, and $N=3$ for a sphere). Then, performing the integration over a circle or a sphere, one can find that
\begin{equation}
\label{eq:05-04-04}
f_\text{max} h \int_{\Omega^R_\text{s}} R^{(i-1)} ~ dV_N^3
\leq
f_\text{max} h \dfrac{4\pi}{i+2} R_0^{(i+N-1)}.
\end{equation}
Since $R_0 \rightarrow 0^+$, the RHS of (\ref{eq:05-04-04}) vanishes as long as $(i+N-1) > 0$, which will always be the case.

\paragraph{Singular surface}
Since the second term of the LHS of (\ref{eq:05-04-05}) vanishes for all the cases encountered in this paper, we want to derive a sufficient condition for the following identity to be true:
\begin{equation}
\label{eq:05-04-06}
\int_{\Omega_\text{ns}} \mathbf{\nabla} \cdot \mathbf{g} ~dV_k^k 
=
\dashint_{\partial \Omega} \mathbf{g} \cdot \hat{n} ~dV_{k-1}^k.
\end{equation}
It can be noticed that, removing $\Omega_\text{s}$ from $\Omega$, the surface of the new volume of integration has been modified in two ways. First, a small part of the surface of $\Omega$ corresponding to the part where $R<R_0$ has been removed. In Figure \ref{fig:div_geom}, it corresponds to the top and bottom surfaces of $\Omega_\text{s}$. Moreover, a new surface $S_a$ corresponding to the interface between $\Omega_\text{ns}$ and $\Omega_\text{s}$ has been created. Applying the divergence theorem and splitting the LHS of (\ref{eq:05-04-06}) into two parts gives
\begin{equation}
\label{eq:05-04-07}
\dashint_{\partial \Omega} \mathbf{g} \cdot \hat{n} ~dV_{k-1}^k
+
\int_{S_a} \mathbf{g} \cdot \hat{n} ~dV_{k-1}^k 
=
\dashint_{\partial \Omega} \mathbf{g} \cdot \hat{n} ~dV_{k-1}^k,
\end{equation}
were the Cauchy integral of the LHS appears due to the fact that the singular zone $R=0$ has been removed from the surface of integration. We can now consider three different cases:

\begin{enumerate}
\item \emph{$\mathbf{g} \cdot \hat{u}_{1\rightarrow3} = 0$ :} \\
Since $\hat{n}$ is orthogonal to the $\hat{u}_{4\rightarrow6}$ directions on $S_a$, if $\mathbf{g}$ is a linear combination these directions, (\ref{eq:05-04-07}) is always true. 

\item \emph{$\mathbf{g} \cdot \hat{u}_{1\rightarrow3} \neq 0$, but  $\lim_{R\rightarrow 0} \big(|\mathbf{g}|R^{(N-1)}\big) = 0$ :} \\
Then, (\ref{eq:05-04-07}) is again verified. Indeed,
\begin{equation}
\label{eq:04-05-08}
\begin{split}
\int_{S_a} \mathbf{g} \cdot \hat{n} ~dV_{k-1}^k 
& \leq \int_{S_a} |\mathbf{g}| ~dV_{k-1}^k \\
& \leq h \int_{\partial \Omega_\text{s}^R} |\mathbf{g}| dV_N^3 \\
& < h g_\text{max} 4\pi R_0^{(N-1)}  \rightarrow 0,
\end{split}
\end{equation}
with $g_\text{max}$ that has been chosen such that $\forall \mathbf{u} \in S_a$, we have $g_\text{max}> |\mathbf{g}|$.

\item \emph{Otherwise :}\\
The case where none of the conditions above is verified and thus (\ref{eq:05-04-07}) is not true is never appearing throughout the paper for the singular zone $R \rightarrow 0$. 

However, in Equation (\ref{eq:06-02-01}), another singular zone appears for $R\neq 0$ and $P_q \rightarrow 0$. Redoing the development of this section for that particular case, it can be seen that the contribution of the small area $P_q< P_{q,0}$, with $P_{q,0} \rightarrow 0^+$, vanishes while the contribution of the additional boundary $P_q = P_{q,0}$ is not negligible and corresponds to the second term of the RHS of (\ref{eq:07-02-01}). For a detailed development, the reader is referred to Equation (5) of \cite{Wilton84} or Equations (39) and (40) of \cite{Jarvenpaa2003}.
\end{enumerate}

\end{appendix}

\begin{IEEEbiography}[{\includegraphics[width=1in,height=1.25in,clip,keepaspectratio]{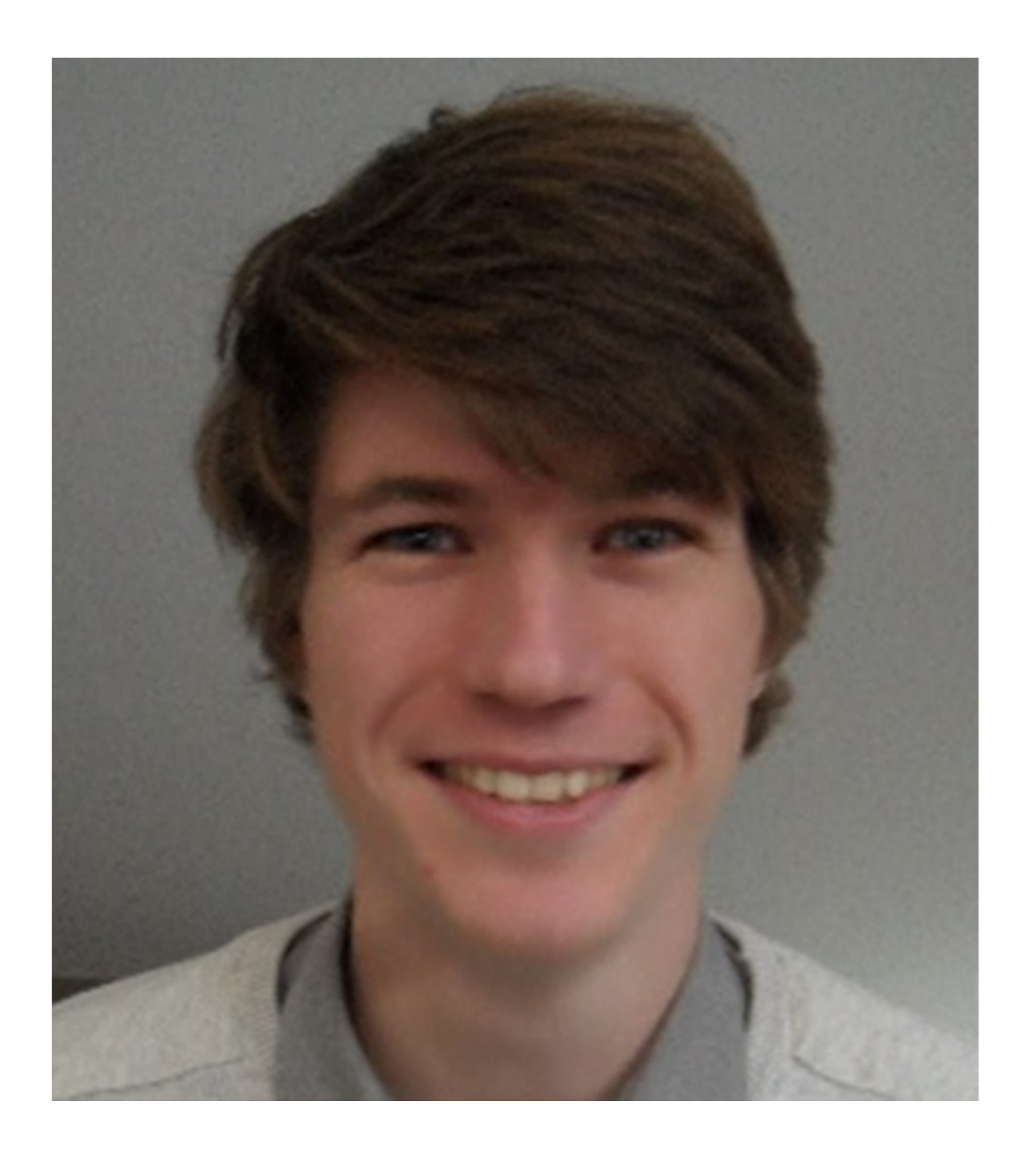}}]{Denis Tihon}
received the B. Sc. and the M. Sc. degrees in Physical Engineering in Universit\'{e} catholique de Louvain (UCL), Louvain-la-Neuve, Belgium, in 2011 and 2013 respectively. Since 2013, he is working as a Teaching Assistant and pursuing a PhD Thesis in the institute of Information and Communication Technologies, Electronics and Applied Mathematics (ICTEAM) in UCL. He is also studying the absorption of partially coherent fields in collaboration with the Cavendish Laboratory, Cambridge, U.K.
His domains of interests include modeling of metamaterials, integral equation based methods and absorption of partially coherent electromagnetic fields.
\end{IEEEbiography}

\begin{IEEEbiography}[{\includegraphics[width=1in,height=1.25in,clip,keepaspectratio]{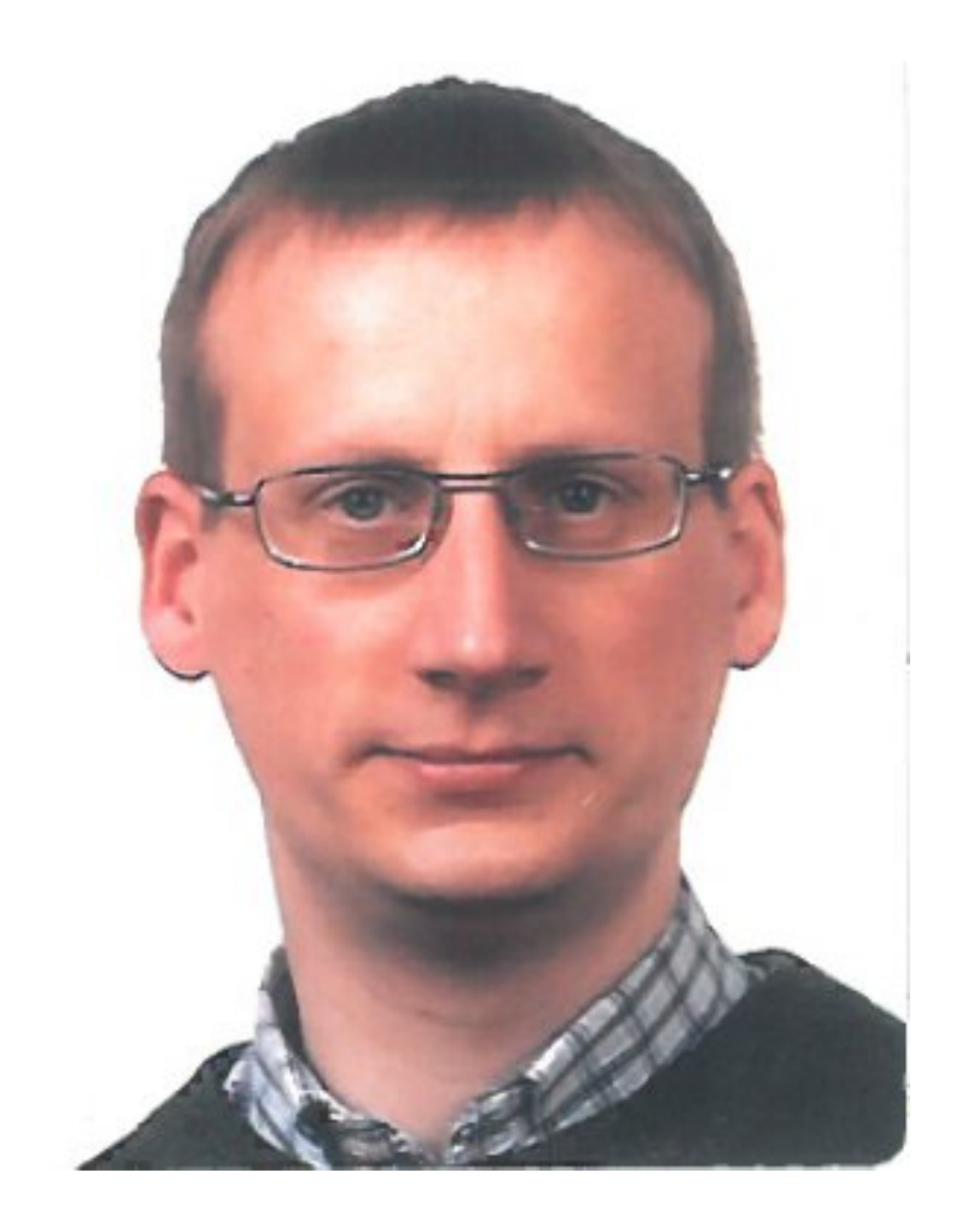}}]{Christophe Craeye} (M'98-SM'11) was born in
Belgium in 1971. He received the Electrical
Engineering and Bachelor in Philosophy degrees
and the Ph.D. degree in applied sciences from the
Universit\'{e} catholique de Louvain (UCL), Louvain-la-Neuve,
Belgium, in 1994 and 1998, respectively.
From 1994 to 1999, he was a Teaching Assistant
with UCL and carried out research on the radar
signature of the sea surface perturbed by rain, in collaboration
with NASA and ESA. From 1999 to 2001,
he was a Postdoctoral Researcher with the Eindhoven
University of Technology, Eindhoven, The Netherlands. His research there was
related to wideband phased arrays devoted to the square kilometer array radio
telescope. In this framework, he was also with the University of Massachusetts,
Amherst, MA, USA, in the Fall of 1999, and was with the Netherlands Institute
for Research in Astronomy, Dwingeloo, The Netherlands, in 2001. In 2002,
he started an antenna research activity at the Universit\'{e} catholique de Louvain,
where he is now a Professor. He was with the Astrophysics and Detectors
Group, University of Cambridge, Cambridge, U.K., from January to August
2011. His research is funded by R\'{e}gion Wallonne, European Commission, ESA,
FNRS, and UCL. His research interests include finite antenna arrays, wideband
antennas, small antennas, metamaterials, and numerical methods for fields in
periodic media, with applications to communication and sensing systems.
Prof. Craeye was an Associate Editor of the \textit{IEEE transactions on antennas and propagation} from 2004 to 2010. He currently serves as an Associate
Editor of the \textit{IEEE Antennas and Wireless Propagation Letters}. In 2009, he was
the recipient of the 2005-2008 Georges Vanderlinden Prize from the Belgian
Royal Academy of Sciences.

\end{IEEEbiography}

\end{document}